\documentclass[aps,preprintnumbers,showpacs,nofootinbib]{revtex4}
\usepackage[english]{babel}
\usepackage{graphicx}
\usepackage{subfig}
\usepackage{hyperref}
\usepackage{dcolumn}
\usepackage{bm}
\newcommand{\newc}{\newcommand}
\newc{\ra}{\rightarrow}
\newc{\lra}{\leftrightarrow}
\newc{\beq}{\begin{equation}}
\newc{\eeq}{\end{equation}}
\newc{\barr}{\begin{eqnarray}}
\newc{\earr}{\end{eqnarray}}

\def\vbf{\mbox{\boldmath $\upsilon$}}
\def\barr{\begin{eqnarray}}
\def\earr{\end{eqnarray}}

\newcommand{\GeV}{\,\textrm{GeV}}

\begin{document}
\rightline{\vbox{\small\hbox{\tt NITS-PHY-2012006} }}
\vskip 1.8 cm

\title{Direct Dark Matter Detection-A spin 3/2 WIMP candidate}
\author{ K. G. Savvidy$^1$ and  J. D. Vergados$^2$ \thanks{Vergados@cc.uoi.gr}}
\affiliation{$^1$ Department of Physics, Nanjing University, Hankou Lu 22, Nanjing, 210098, China\\$^2$ \it Theoretical Physics Division, University
of Ioannina, Ioannina, GR 451 10, Greece}
\begin{abstract} 
We consider a Dark Matter candidate particle of spin 3/2 with neutrino-like Standard Model strength interactions.  In the Majorana case, the particle can account for all of the Dark Matter for a range of masses between 70-160 GeV, depending on the strength of the Higgs couplings. The elementary spin-dependent cross section on the nucleon is calculated to be $1.7 \times 10^{-2}\mbox{ pb}$, and does not depend on the mass or any additional parameters.  The Dirac case  is excluded by the very large coherent cross section. The amplitude at the nuclear level is of the purely isovector form. We make detailed predictions for differential and total rates of scattering on a variety of nuclear targets of interest to the current direct detection experiments. For heavy targets the annual modulation amplitude is predicted to be negative, which may help to determine the mass of the WIMP if and when data become available.
\end{abstract} 
\date{\today}
\maketitle

\section{Introduction}
The combined MAXIMA-1 \cite{MAXIMA-1}, BOOMERANG \cite{BOOMERANG},
DASI \cite{DASI} and COBE/DMR Cosmic Microwave Background (CMB)
observations \cite{COBE} imply that the Universe is flat
\cite{flat01}
and that most of the matter in
the Universe is Dark \cite{SPERGEL},  i.e. exotic. These results have been confirmed and improved
by the recent WMAP data \cite{WMAP06}. Combining 
the data of these quite precise measurements one finds:
$$\Omega_b=0.0456 \pm 0.0015, \quad \Omega _{\mbox{{\tiny CDM}}}=0.228 \pm 0.013 , \quad \Omega_{\Lambda}= 0.726 \pm 0.015~.$$
Since any ``invisible" non exotic component cannot possibly exceed $40\%$ of the above $ \Omega _{\mbox{{\tiny CDM}}}$
~\cite {Benne}, exotic (non baryonic) matter is required and there is room for cold dark matter candidates or WIMPs (Weakly Interacting Massive Particles).

Even though there exists firm indirect evidence for a halo of dark matter
in galaxies from the
observed rotational curves, see e.g. the review \cite{UK01}, it is essential to directly
detect
such matter.
Until dark matter is actually detected, we shall not be able to
exclude the possibility that the rotation curves result from a
modification of the laws of nature as we currently view them.  This makes it imperative that we
invest a
maximum effort in attempting to directly detect dark matter in the laboratory. Furthermore such a direct detection will also
unravel the nature of the constituents of dark matter.
The possibility of such detection, however, depends on the nature of the dark matter
constituents and their interactions.

Since the WIMP's are  expected to be
extremely non relativistic, with average kinetic energy $\langle T\rangle  \approx
50 \ {\rm keV} (m_{\mbox{{\tiny WIMP}}}/ 100 \ {\rm GeV} )$, they are not likely to excite the nucleus, even if they are quite massive $m_{\mbox{{\tiny WIMP}}} > 100$ GeV.
Therefore they can be directly detected mainly via the recoiling of a nucleus
(A,Z) in elastic scattering. Computing the event rate for such a process can
requires the following ingredients:
\\i) An effective Lagrangian at the elementary particle (quark)
level obtained in the framework of the prevailing particle theory.  The most popular scenario is in the context of supersymmetry. Here  the dark matter candidate is the
LSP (Lightest Supersymmetric Particle) \cite{ref2a,ref2b,ref2c,ref2,ELLROSZ,Gomez,ELLFOR}.
In this case the effective Lagrangian is constructed as described,
e.g., in Refs.~\cite{ref2a,ref2b,ref2c,ref2,ELLROSZ,Gomez,ELLFOR,JDV96,JDV06a}. At least till supersymmetry is discovered in the laboratory (LHC) other approaches are also possible, such as, e.g., technibaryon \cite{Nussinov92,GKS06}, mirror matter\cite{FLV72,Foot11}, Kaluza-Klein models with universal extra dimensions\cite{ST02a,OikVerMou} 
\\ii) A well defined procedure for transforming the amplitude thus
obtained, using the previous effective Lagrangian, from the quark to
the nucleon level. To achieve this one needs a quark model for the nucleon, see e.g.  \cite{JDV06a,Dree00,Dree,Chen}. This step is particularly important in supersymmetry or other models dominated by a scalar interaction (intermediate Higgs etc), since, then, the elementary amplitude becomes proportional to the quark mass and the content of the nucleon in quarks other than $u$ and $d$  becomes very important.
\\iii) knowledge of the relevant nuclear matrix elements
\cite{Ress,DIVA00}, obtained with as reliable as possible many
body nuclear wave functions, 
\\iv) knowledge of the WIMP density in our vicinity and its velocity distribution.

From steps i) and ii) one obtains the nucleon cross sections. These can also be extracted from experimental event rates, if and when such data become available. Typically, experiments obtain upper limits on the event rates, so one can get exclusion plots on the nucleon cross sections as a function of the WIMP mass. The extracted cross sections are subject to uncertainty of the nuclear and astrophysical inputs from steps iii)-iv).

In the standard nuclear recoil experiments, first proposed more than 30 years ago \cite{GOODWIT}, one has to face the problem that the reaction of interest does not have a characteristic feature to distinguish it
from the background. So for the expected low counting rates the background is
a formidable problem. Some special features of the WIMP-nuclear interaction can be exploited to reduce the background problems. Such are:

i) the modulation effect\cite{Druck},\cite{PSS88,GS93,RBERNABEI95,LS96,ABRIOLA98,HASENBALG98,JDV03,GREEN04,SFG06}
ii) backward-forward asymmetry expected in directional experiments, i.e. experiments in which the direction of the recoiling nucleus is also observed \cite{SPERGEL88}\cite{DRIFT,SHIMIZU03,KUDRY04,DRIFT2,GREEN05,Green06,KRAUSS,KRAUSS01,Alenazi08,Creswick010,Lisanti09,Giometal11}.

 In connection with nuclear structure aspects, in a series of calculations, e.g. in \cite{JDV03,JDV04,VF07} and references therein, it has been shown that for the coherent contribution, due to the scalar interaction, the inclusion of the nuclear form factor is important, especially in the case of relatively heavy targets. They also showed that the nuclear spin cross sections  are characterized by a single, i.e. essentially isospin independent, structure function and two static spin values, one for the  proton and one for the neutron, which depend on the target.
 
  As we have already mentioned, an essential ingredient in direct WIMP detection is the WIMP density in our vicinity and, especially, the WIMP velocity distribution. Some of the calculations have considered various forms of phenomenological non symmetric velocity distributions  \cite{DRIFT2,GREEN04,GREEN05} and some of them even more exotic dark matter flows like
the late infall of dark matter into  the galaxy, i.e caustic rings
 \cite{SIKIVI1,SIKIVI2,Verg01,Green,Gelmini}, dark matter orbiting the
 Sun \cite{KRAUSS} and Sagittarius dark matter \cite{GREEN02}.  We will employ here the standard Maxwell-Boltzmann (M-B)
distribution for the WIMPs of our galaxy and we will not be concerned with other  distributions
\cite{VEROW06,JDV09,TETRVER06,VSH08}.
 
  In addition to computing the time averaged rates, these works studied the modulation effect. They showed that in the standard recoil experiments the modulation amplitude  in the total rate may change sign for large reduced mass, i.e. heavy WIMPS and large A. This may be exploited as an additional discriminant to deduce the mass of the WIMP from the data, if and when they become available.
There exists a plethora of direct dark matter experiments  with the task of detecting  WIMP event rates   for a variety of targets,  such as those employed in XENON10 \cite{XENON10}, XENON100 \cite{XENON100.11}, XMASS \cite{XMASS09}, ZEPLIN \cite{ZEPLIN11}, PANDA-X \cite{PANDAX11}, LUX \cite{LUX11}, CDMS \cite{CDMS05}, CoGENT \cite{CoGeNT11}, EDELWEISS \cite{EDELWEISS11}, DAMA \cite{DAMA1,DAMA11}, KIMS \cite{KIMS07} and PICASSO \cite{PICASSO09,PICASSO11}.

In the present paper we will employ these well-understood ingredients and consider  another attractive candidate, namely a spin 3/2 particle suggested in \cite{Savvidy:2006yj, Savvidy:2005vm}, which  has some interesting properties and satisfies all the required constraints. Our detailed calculations of relic abundance in Section \ref{sec:abundance} show that the candidate can account for Dark Matter in the mass range between 70-160 GeV, which is mildly dependent on two additional parameters. This particle couples to the nucleon via Z-exchange. Therefore it can lead to large spin independent and spin dependent cross sections depending on whether it is a Dirac or Majorana particle. In both cases the required cross-sections and differential and total rates are calculated with only one unknown parameter, namely the mass of the particle. The former  leads to coherence of all the neutrons in the target, and can be already excluded by presently available data. If, however, it is a Majorana fermion, the time component of the neutral current contributes only at order $\beta^2 \approx 10^{-6}$ and the coherence is suppressed. In this case   the direct detection process is dominated by the isovector, purely spin cross section. For the nucleon we calculate this cross-section in Section \ref{sec:nuc}, with essentially no model uncertainty and obtain the value $\sigma_N^{\mbox{\tiny{spin}}} = 1.7 \times 10^{-2}\mbox{ pb}$. We present the differential and total direct detection rates for a variety of target nuclei, for both a light and a heavy WIMP in Section \ref{sec:rates}. We conclude that the model, while being subject to the usual astro and nuclear physics uncertainties but in principle no adjustable parameters, can serve as a benchmark for experiments with nuclei which are sensitive to spin, including those with Xenon, and that one can be optimistic that the required sensitivities will be achieved in the near future.

\section{A spin 3/2 particle as a dark matter candidate.}
As we have mentioned in the introduction, the direct detection of dark matter is central to physics and cosmology. Among other things, it will establish the nature of the dark matter constituents and will discriminate among the various particle models. We will consider here a neutral spin 3/2 particle  \cite{Savvidy:2006yj, Savvidy:2005vm} ,  which  was proposed as part of an extension of the electroweak sector of the SM with gauge bosons of spin 2 and matter of spin 3/2. 

It is assumed  that a minimum extension of the standard model (SM) takes place so that the left handed component is put in a isodoublet of the SU(2)$_L$ and the right handed partners in isosinglets,
\beq
\label{mult}
X_L=\left ( \begin{array}{c}\chi_{_L}\\ l^-_{_L} \end{array} \right ) ~\mbox{ with }Y=-1~, \quad \chi_{_R} \mbox{ with } Y=0~, \quad l^-_{_R} \mbox{ with } Y=-2 ~.
\eeq 
These spin 3/2 particles should be described by a Lorentz structure transforming as a vector-spinor; for brevity we have suppressed these indices. Thus, except for the different spin, it has the quantum numbers of a lepton doublet: $\chi$ is neutral and $l$ has unit negative charge according to the hypercharge assignments above. 

\begin{figure}[!t]
\begin{center}
\includegraphics[scale=0.7]{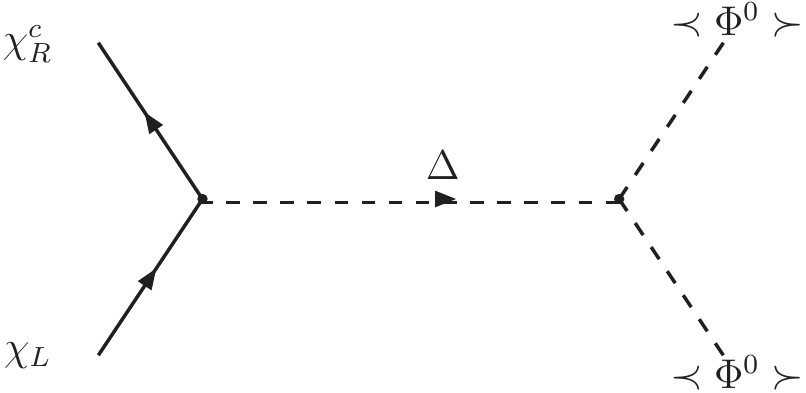}
 \caption{ The tree level contribution to the neutrino mass mediated by  an isotriplet scalar.}
 \label{Fig:tripletDelta}
\end{center}
\end{figure}

This particle possesses Yukawa couplings like every other fermion:
\beq
{\cal L}_{\chi}=f_{\chi} \, \left (\bar{\chi}_{_L},\bar{l}^-_{_L} \right )\, \left ( \begin{array}{c}\phi^0\\ \phi^- \end{array}\right ) \, \chi_{_R},
\quad {\cal L}_{\chi}=f_{\chi}\, \left (\bar{\chi}_{_L},\bar{l}^-_{_L} \right ) \, \left ( \begin{array}{c}\phi^+\\ \phi^{*0} \end{array}\right ) l^-_{_R}
\eeq   
and it can acquire a Dirac mass like every other fermion, i.e. spontaneously when the isodoublet  acquires a vacuum expectation value
\beq
m^{D}_{\chi}=f_{\chi}\, \frac{\upsilon}{\sqrt{2}}
\eeq
 The neutral component may also acquire a Majorana mass like the neutrino, in a see saw like manner. This can be achieved directly in the presence of an isotriplet $\Delta$ of Higgs scalars \cite{MohEtal07,MohSmyr06} whose charge decomposition is $\delta^{--},\delta^-,\delta^0$. \\Then this leads to the coupling:
$$
	{\cal L}_{\Delta,\chi}=	h_{\Delta,\chi} ~ (\bar{\chi}_{_L}\bar{l}^-_{_L}) \, 
		\left(
\begin{array}{lr}\delta^-&-\delta^0\\
                    \delta^{--}& \delta^-\\                                  
\end{array}                      
\right )\left(
\begin{array}{c}l^{-c}_{ R}\\
                    -\chi^{c}_{ R}\\                  
\end{array}                      
\right )$$
This after the isotriplet acquires a vacuum expectation value becomes
$${\cal L}_{\Delta,\chi}=	h_{\Delta,\chi} \left (\bar{\chi}_{_L}\bar{l}^-_{_L} \right )
		\left(
\begin{array}{lr}0&-v_{\Delta}/\sqrt{2}\\
                    0&0                                  
\end{array}                      
\right )
\left (
\begin{array}{c}l^{-c}_{ R}\\
                    -\chi^{c}_{ R}                                 
\end{array}                      
\right )$$
yielding the neutrino Majorana mass:
\beq
m^{M}_{\chi}=h_{\Delta,\chi}\frac{v_{\Delta}}{\sqrt{2}}
\label{Eq:tripletmass}
\eeq
Alternatively  the Majorana mass matrix can be obtained assuming that the isotriplet $\Delta$ possesses a  cubic coupling $\mu_{\Delta} $ with  two standard Higgs doublets \cite{MagWet80,LazShafWet81,MohSen81,Ver86} (see Fig. \ref{Fig:tripletDelta}). This  effective cubic coupling yields a Majorana mass term through the vacuum expectation value of the isodoublet.
The effective Majorana neutrino mass is
obtained from Eq. (\ref{Eq:tripletmass}) via the substitution
 \beq
 \frac{v_{\Delta}}{\sqrt{2}}\rightarrow \frac{v^2}{2}\frac{\mu_{\Delta}}{m^2_{\Delta}},
 \label{dt}
 \eeq
 where $v/\sqrt{2}$ is the vacuum expectation value of the standard Higgs doublet, $m_{\Delta}$ is the mass of $\delta^0$.
   The cubic coupling may be thought of arising out of a quartic coupling $\lambda$ in the presence of an isosinglet acqiring a vacuum expectation value $\upsilon_R$, in which case $\mu_{\Delta}=\lambda \upsilon_R$.
   \\We note here that, if the isotriplet $\Delta$ possesses a  cubic coupling $\mu_{\Delta} $ with  two standard Higgs doublets, then this mechanism could lead to WIMP nucleus scattering via Higgs exchange, with one of the Higgs acquiring a vacuum expectation value. When it comes to the hadronic vertex this contribution is analogous to the LSP (light supersymmetric particle of supersymmetry). In any case, even if the  situation is otherwise favorable, i.e. $\mu_{\Delta}\times \upsilon \approx m^2_{\Delta}$,
  this mechanism is going to be suppressed compared to the Z-exchange, due to the smallness of the  mass  of the dominant quarks present in the nucleus.  We will not elaborate further on this mechanism.   

If this model is to account for Dark Matter, the Yukawa couplings should be chosen such that the neutral particle is lighter than the charged partner. In this case one may ask whether the particle is stable with respect to decay to SM particles. Fortunately, there is a natural explanation for stability of the lightest BSM particle. According to the quantum number assignment, there may exist a vertex for the decay of the neutral $\chi$ into a  $W^+$  and an electron $e^-$  (and in the Majorana case also a $W^-$ and a positron $e^+$). Owing to the Lorentz structure, such vertex cannot be of the minimal type, and in fact in the model described in \cite{Savvidy:2005vm} is completely prohibited even if any possible loop effective contributions are included. Nevertheless, one may write down a dimension-5 operator such as 
\beq
L_{\tiny{int}} = \frac{1}{M_*} \,  \bar{e} \,  \gamma_\mu \,  \chi_\nu ~ (q^\mu \,  W^+_\nu - q^\nu \, W^+_\mu) + \textrm{h.c.}
\eeq
The interaction is therefore suppressed by some high mass scale $M_*$, which may be identified with the Planck scale. In this case the decay width is estimated to be $\Gamma \approx \alpha m_\chi \,(\frac{m_e}{M_{pl}})^2 \approx 10^{-45} \GeV$ which is sufficient to make the particle long lived on the cosmological scale.

\section{Relic abundance}
\label{sec:abundance}
 Some years ago, Lee and Weinberg \cite{Lee:1977ua} obtained a lower bound on the mass of a stable neutrino-like Dirac massive particle from the condition that it should not over-close the universe. With the couplings precisely those of the SM neutrino, and modern value for dark matter abundance, $\Omega_{DM} h^2 =  0.1099 \pm 0.0062$ one obtains the estimate $M_\nu \approx 10 \GeV$ with the cross section for annihilation at rest into a light fermion 
 \beq
 \label{eq:LW}
\sigma\, v =  \frac {G_F^2 \, M_{\nu}^2}{2\pi }  \approx 1 pb . 
 \eeq
The whole scenario became disfavored as such particle was not seen experimentally, and moreover LEP limited the number of neutrinos lighter than $\approx 45 \GeV$ to just three. Further, it was convincingly demonstrated by Kainulainen et al \cite{Enqvist:1988we} that although the annihilation crosssection drops somewhat after the Z peak, nevertheless the relic abundance of such a particle does not reach the value necessary to overclose the universe. The dominant annihilation channel was found by them to be into $W^+, W^-$ pairs for particles more massive than about 100 GeV. Moreover, the crosssection keeps growing for high values of the mass, in part because of a large and growing coupling to the Higgs. This feature was criticised in \cite{Griest:1989wd} on very general grounds, arguing that the thermally averaged cross section should drop like $\alpha/M_{\chi}^2$ in all channels. Nevertheless, the calculations of   Kainulainen et al show that it comes close to being able to account for the Dark Matter just before this problematic channel opens up, $\Omega <\approx 0.1$ for  $M_{\chi} \approx 80$GeV.  

  It is also interesting to ask whether the picture is different if this particle is assumed to be Majorana. In this case the annihilation crosssection to light fermions is suppressed at rest due to the vanishing of the vector current for a majorana particle \cite{Goldberg:1983nd, Kolb:1985nn}; we find that the annihilation into $W^+, W^-$ pairs is also vanishing at rest. The resulting abundance is somewhat higher than in the Dirac case but does not reach the value necessary to account for all of the Dark Matter above the Z peak. We have calculated the abundance for a range of parameters in the case of our spin 3/2 candidate, and found that the abundance may equal the presently well-measured Dark Matter abundance for a range of masses, $M_{\chi} \approx 70-110 \GeV$. 
\begin{figure} [htdp] 
\subfloat[majorana, spin 1/2]{\includegraphics[scale=0.4]{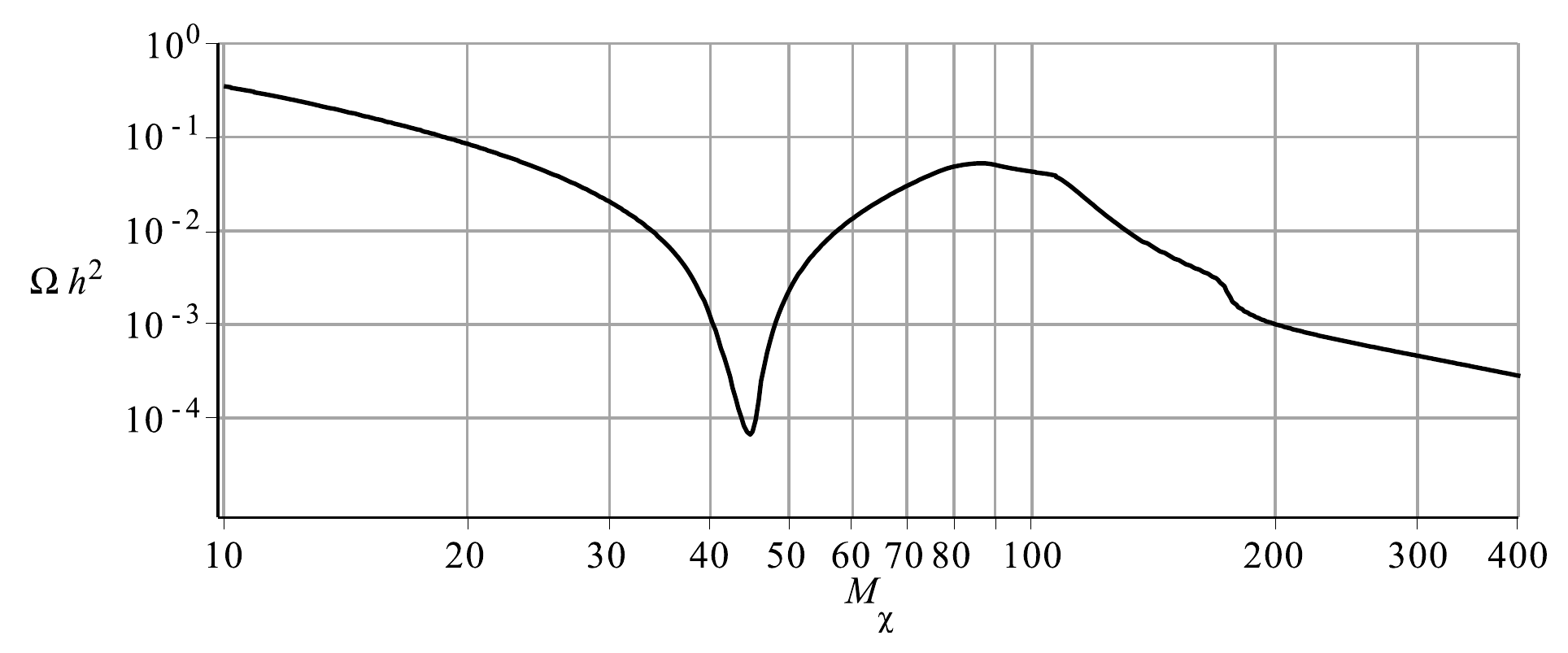}}
\subfloat[majorana, spin 3/2]{\includegraphics[scale=0.4]{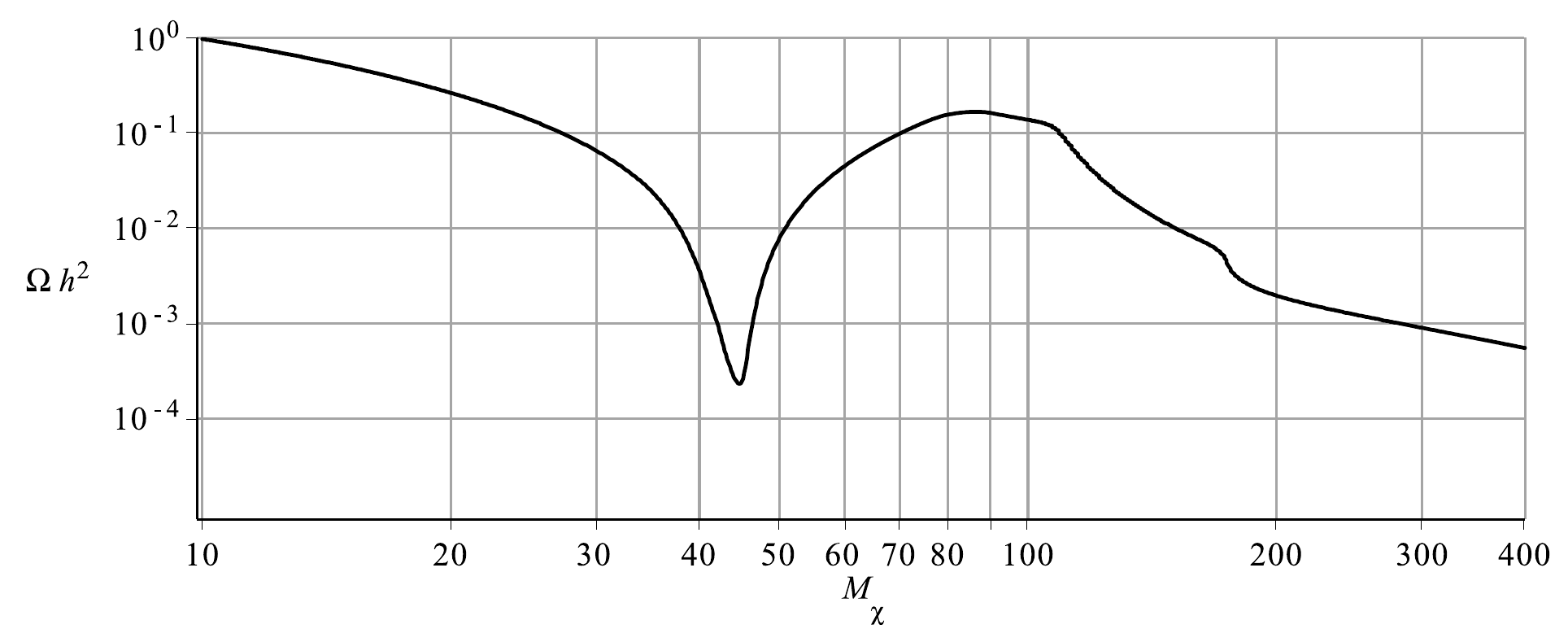}}
\caption{{\small The relic abundance as a function of the mass of the candidate particle for some representative value of the charged lepton mass and no doublet-triplet mixing. Horizontal axis: mass of the DM particle in GeV. Vertical: abundance, $\Omega_{DM} h^2$. The model was implemented and abundances were calculated using Micromegas \cite{Belanger}.}
\label{fig:LWplot}}
\end{figure}
  
  The relic abundance depends mildly on the mass of the charged partner $l^-$ (eq. \ref{mult}). It must be heavier than our neutral candidate, and highest abundances are achieved when it is 2 to 4 times heavier than the neutral particle: if it is any lighter then co-annihilation effects reduce the abundance, and if it is much heavier there is less cancellations due to the $W^+, W^-$ t-channel diagram, also reducing the abundance. In addition the precise value depends also on the amount of Higgs doublet-triplet mixing which determines the relative contribution to the mass of eq.(\ref{Eq:tripletmass}) vs eq (\ref{dt}).

\begin{figure} [htdp] 
\includegraphics[scale=0.6]{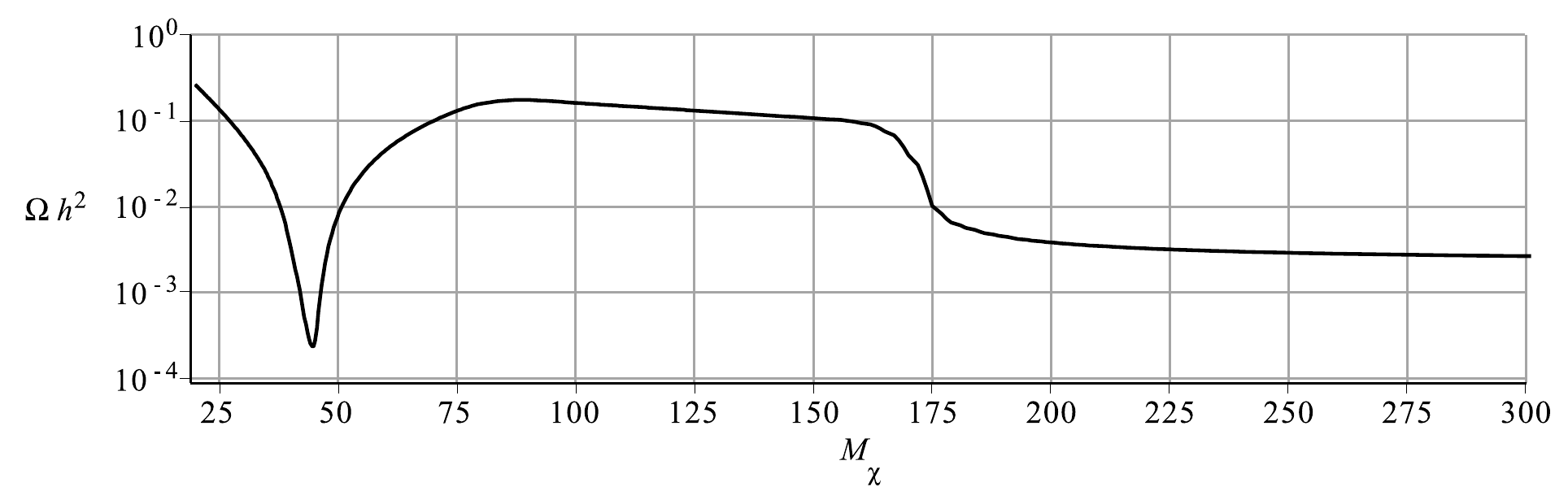}
\caption{{\small The abundance in the case that the Higgs coupling to Z is substantially smaller than the tree-level SM value. Otherwise as in Fig. \ref{fig:LWplot}.}}
\label{fig:nohiggs}
\end{figure}

With the Higgs mass of around 125 GeV, the Z,h channel opens up above $M_{\chi} = 110 \GeV$ and in fact dominates the annihilation crosssection at high values of $M_{\chi}$. Fortunately, as of the writing of this paper, we now know the value of the Higgs mass, but the couplings of this particle have not been yet verified experimentally. At LHC, the four lepton decay channel of the Higgs probes the ZZh coupling. We have used the Standard Model couplings, but in some extensions of the SM the ZZh coupling is modified significantly; this may extend somewhat the allowed range of masses to 160 GeV (see Fig. \ref{fig:nohiggs}).

\section{The direct detection formalism}
\label{sec:nuc}
\begin{figure}
\includegraphics[width=0.4\textwidth]{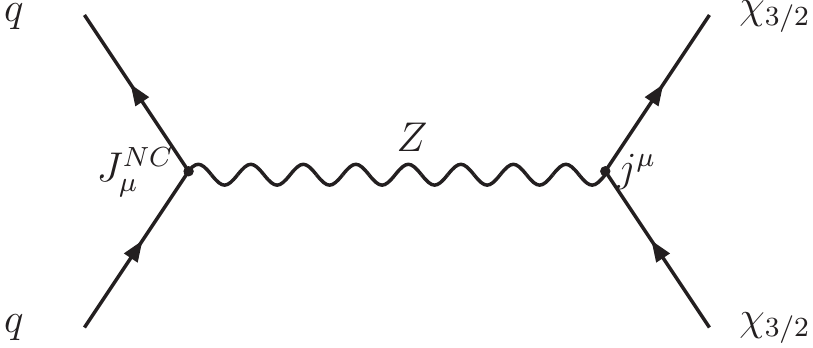}
\caption{ The  $\chi_{3/2}$ interaction with quarks in the nucleus is mediated by Z-exchange.
 \label{LSPZ} }
\end{figure}

 The main advantages of our WIMP candidate are:
\begin{itemize}
\item It can interact with the quarks via exchange of Z, a relatively light particle with well known interactions (Fig. \ref{LSPZ}). 
\item It can lead to both coherent and spin dependent modes.\\
The rate for the coherent mode is proportional to $\sigma_N^{\mbox{\tiny{coh}}}N^2$, i.e. to the square of the neutron number of the target.\\
 The spin dependent cross section is purely isovector, i.e. the spin dependent rate is proportional to $\sigma_N^{\mbox{\tiny{spin}}}(\Omega_p-\Omega_n)^2$, i.e. on one particle model parameter, $\sigma_N^{\mbox{\tiny{spin}}}$. $\Omega_p$, $\Omega_n$ are the proton and neutron components of the nuclear spin matrix elements which are known for most targets of experimental interest. The elementary nucleon cross sections, $\sigma_N^{\mbox{\tiny{coh}}}$ and $\sigma_N^{\mbox{\tiny{spin}}}$ are going to be calculated in this work.
\end{itemize}  

In the standard weak interaction  the neutral current contact interaction takes the form:
\beq
H=\frac{G_F}{\sqrt{2}} j^{\mu}_L J^{NC}_{\mu}
\eeq
  $J^{NC}_{\mu}$ is the  hadronic current at the nucleon level, which in the non relativistic limit is of the form:
\beq
J^{NC}_{\mu}=\frac{1}{2}\left ((-1+4 \sin^2{\theta_W}), \,g_A \sigma\right )\approx\left(0,\,\frac{g_A}{2}\sigma \right ), \mbox{ for the proton}
\eeq
i.e. the time component almost vanishes. Also
\beq
J^{NC}_{\mu}=\left ((-\frac{1}{2}, \,-\frac{g_A}{2} \sigma\right ), \mbox{ for the neutron}
\eeq
Note that the space component of the neutron is opposite to that of the proton.

The  current $j^{\mu}$ is associated with the WIMP,  given in \cite{SAVVIDI11}, except that here we also have chiral interactions,
\beq
j^\mu =  \bar{\chi}^\nu \, \Gamma^\mu_{\nu\rho} \, ( \Lambda_V+\gamma_5 \, \Lambda_A) \, \chi^\rho 
\eeq
where $\Gamma^\mu_{\nu\rho}$ can be replaced by $\gamma^\mu \, \eta_{\nu\rho}$ in this particular case.
%
The calculated currents can be cast into a form similar to that of the nucleon with the use of the standard quantum mechanical spin 3/2 angular momentum operators $\vec{L}_{3/2}$,
\beq
j^\mu_L = \left( \Lambda_V \, , ~ \frac{2}{3}\, \vec{L}_{3/2} \, \Lambda_A\right ) 
\eeq
here $\Lambda_V$ and $\Lambda_A$ are the vector and axial couplings of the particle.

At this point we encounter the following possibilities:

 \begin{enumerate}
 \item The coherent, spin independent (SI) rate.\\
We will first consider the case that the spin 3/2 particle is a Dirac particle. In this case, it is natural to take the neutral current to be of the purely left handed form, with $\Lambda_V = -\Lambda_A =1 $. 
Assuming that the WIMP is a Dirac type particle and the vector coupling does not vanish, since $\Lambda_V=1$, the nuclear cross section is dominated by its coherent component and is mostly due to the neutron. The coherent part of the neutron cross section can be cast in the form
\beq
 \sigma_N^{\mbox{\tiny{coh}}}= \frac{G^2_F m^2_N}{2 \, \pi}~ ~ \frac{1}{8}\sum_{{M_i,M_f,}{\\m_i,m_f}}
 \big|{\cal M}_0\left (M_i,m_i\rightarrow M_f, m_f\right )\big|^2
\eeq
where $M_i,m_i$ are the oncoming spin projections of the oncoming 3/2 WIMP and nucleon respectively, while $M_f,m_f$ are the corresponding quantities for the outgoing particles. ${\cal M}$ is the invariant amplitude. The last factor of 1/8 arises  from the averaging over the initial spin states of the WIMP and the nucleon.\\
 The spin independent amplitude  is diagonal and takes the form:
\beq
{\cal M}_0\left (M_i,m_i\rightarrow M_f, m_2\right )=-\frac{1}{2} \, \delta_{M_i,M_f}\delta_{m_i,m_f}
\eeq
Averaging over the nucleon spins we get:
\beq
\frac{1}{8} ~ \sum_{M_i,M_f,m_i,m_f}\big| {\cal M}_0\left (M_i,m_i\rightarrow M_f, m_f\right ) \big|^2=\frac{1}{8} ~ 4 \times 2 ~ \left (\frac{1}{2}\right )^2  = 1/4
\eeq
\beq
 \sigma_N^{\mbox{\tiny{coh}}}=\frac{G^2_F m^2_N}{2 \, \pi} ~ \frac{1}{4}~  = \frac{G^2_F m^2_N}{8 \, \pi} 
\eeq 
Thus in our model: 
\beq
\sigma_N^{\mbox{\tiny{coh}}}\approx 1.66 \times 10^{-3} \mbox{pb}
\eeq

\item The spin dependent (SD) cross section.\\
Next we will consider the case that this particle is of Majorana nature, i.e. the particle coincides with its antiparticle. In this case the time component of the WIMP current vanishes in the non relativistic limit, since effectively $\Lambda_V=0$. The cross section is then of the purely isovector form, and is identical for the neutron and the proton due to their equal axial coupling:
\beq
\sigma_N^{\mbox{\tiny{spin}}}=\frac{G^2_F \, m^2_N}{2 \, \pi} ~ g^2_A\, \frac{1}{8}\sum_{M_i,M_f,m_i,m_f} 
\big| {\cal M}_1\left (M_i,m_i\rightarrow M_f, m_f\right )\big|^2
\eeq
where ${\cal M}_1\left (M_i,m_i\rightarrow M_f, m_f\right )$ is given in the table \ref{tab:amplitudes}.
\begin{table}[!t]
\caption{The various matrix elements ${\cal M}$ of the axial current interaction. $M_i$, $M_f$ stand for the spin projections of the initial and final spin 3/2 WIMP, while $m_i$, $m_f$ are corresponding spin projections of the nucleon. The nucleon isospin projection was taken to be +1/2.} 
\label{tab:amplitudes}
\begin{center}
\begin{tabular}{||l|c|c|c||c||}
\hline
$M_i$&$M_f$&$m_i$&$m_f$&${\cal M}$\\
\hline
3/2&3/2&$\pm$ 1/2&$\pm$ 1/2 &$\pm$1/2\\
-3/2&-3/2&$\pm$ 1/2&$\pm$ 1/2 &$\pm$1/2\\
1/2&1/2&$\pm$ 1/2&$\pm$ 1/2 &$\pm$1/6\\
-1/2&-1/2&$\pm$ 1/2&$\pm$ 1/2 &$\pm$1/6\\
3/2&1/2&-1/2& 1/2 &1/$\sqrt{3}$\\
-3/2&-1/2& 1/2& -1/2 &1/$\sqrt{3}$\\
1/2&3/2&- 1/2& 1/2 &1/$\sqrt{3}$\\
-1/2&-3/2& 1/2& -1/2 &1/$\sqrt{3}$\\
1/2&-1/2&- 1/2& 1/2 &2/3\\
-1/2&1/2& 1/2& -1/2 &2/3\\
\hline
\end{tabular}
\end{center}
\end{table}
A straightforward  summation over the nucleon spins  results in:
\[
 \frac{1}{8} \, \left (4 \times\frac{1}{4} + 4\times\frac{1}{36} + 4\times\frac{1}{3} + 2\times\frac{4}{9}\right )=\frac{5}{12}
\]
The nucleon cross section is then
\beq
\sigma_N^{\mbox{\tiny{spin}}}=
\frac{G^2_F m^2_N}{2 \, \pi} ~ g^2_A \frac{5}{12}
\eeq
Thus in our model, and with $g_A\approx 1.24$,
\beq
\sigma_N^{\mbox{\tiny{spin}}}(\mbox{(Dirac)}\approx 4.25 \times 10^{-3}\mbox{ pb},\quad\sigma_N^{\mbox{\tiny{spin}}}(\mbox{(Majorana)}\approx 1.70 \times 10^{-2}\mbox{ pb}
\eeq
\end{enumerate}
i.e. in the Dirac case the spin-dependent cross section is not of interest to us, since it is sub dominant for all targets of interest: the spin independent mechanism dominates due to coherence. The Majorana case in interesting since in this case there is no competing coherent mode. Note that in our model the spin dependent cross section does not depend on the heavy-quark content of the nucleon.
We note that this large cross section in the case of the Majorana WIMP is close to the current limits extracted from the recent KIMS data \cite{KIMS12} and   the earlier ZEPLIN data \cite{ZEPLIN09}  
$$2\times 10^{-2}\mbox{ KIMS CsI(T$\ell$)},\quad 1.8 \times 10^{-2}\mbox{ ZEPPLIN-III collaboration}$$
for the dominant cross section relevant to their target.

It is now straightforward to  find the cross sections at the nuclear level. In the first case, the cross-section is coherent and scales quadratically with the number of the neutrons:
\beq
\sigma^{\mbox{\tiny coh}}_{\mbox{\tiny nuclear}}=N^2 \, \sigma_N^{\mbox{\tiny{coh}}} \, F^2(q)
\eeq
where ~$F(q)$ is the nuclear form factor and in the second case it is pure spin and is of the isovector form:
\beq
\sigma^{\mbox{\tiny spin}}_{\mbox{\tiny nuclear}}=\frac{1}{3} \, \left( \Omega_p-\Omega_n \right)^2 \, \sigma_N^{\mbox{\tiny{spin}}} \, F_{11}(q)
\eeq
where$ F_{11}(q)$ is the spin response function, normalized so that it is unity at zero momentum transfer, and $\Omega_p$, $\Omega_n$ are the static values of the nuclear proton and neutron spin components for the target nucleus, normalized so that they become equal to 3 for the free nucleon.

\section{The formalism for the WIMP-nucleus differential event rate}
The formalism adopted in this work is well known (see e.g. the recent reviews \cite{JDV06a,VerMou11}). So we will briefly discuss its essential elements here.
Before calculating the direct detection event rate, we will first deal with the WIMP velocity distribution. To this end we will follow the steps:
\begin{itemize}
\item one starts with such distribution in the Galactic frame.
\item one transforms to the local coordinate system:
\beq
{\bf y} \rightarrow {\bf y}+{\hat\upsilon}_s+\delta \left (\sin{\alpha}{\hat x}-\cos{\alpha}\cos{\gamma}{\hat y}+\cos{\alpha}\sin{\gamma} {\hat \upsilon}_s\right ) ,\quad y=\frac{\upsilon}{\upsilon_0}
\label{Eq:vlocal}
\eeq
with $\gamma\approx \pi/6$. $ \hat {\upsilon}_s$ a unit vector in the Sun's direction of motion, ${\hat x}$  a unit vector radially out of the galaxy in our position and  $\hat{y}={\hat \upsilon}_s\times \hat{x}$. The last term in the first expression of Eq. (\ref{Eq:vlocal}) corresponds to the motion of the Earth around the Sun with $\delta$ being the ratio of modulus of the Earth's velocity around the Sun divided by the Sun's velocity around the center of the Galaxy, i.e.  $\upsilon_0\approx 220$km/s and $\delta\approx0.135$. The above formula assumes that the motion  of both the Sun around the Galaxy and of the Earth around the Sun are uniformly circular. The exact orbits are, of course, more complicated \cite{GREEN04,LANG99}, but such deviations are not expected to significantly modify our results. In Eq. (\ref{Eq:vlocal}) $\alpha$ is the phase of the Earth ($\alpha=0$ around June 3nd)\footnote{One could, of course, make the time dependence of the rates due to the motion of the Earth more explicit by writing $\alpha \approx(6/5)\pi\left (2 (t/T)-1 \right )$, where $t/T$ is the fraction of the year.}.
\item One integrates  the velocity distribution over the angles and the result is multiplied by the  velocity $\upsilon$ due to the WIMP flux.
\item The result is integrated from a minimum value $\upsilon_{min}$ to the maximum allowed velocity  $\upsilon_{max}$. In general, the escape velocity $ \upsilon_{esc}$ in our galaxy is estimated to be in the range  550km/s$\le\upsilon_{esc}\le650$km/s. In our calculations we assumed for the M-B distribution  $\upsilon_{esc}\approx620$km/s,  even though the value\cite{spergel12} of 550 km/s, which results from an analysis of data from the RAVE survey \cite{RAVE06}, would have been more appropriate. The obtained results are not sensitive to this value. $\upsilon_{min}$ is a suitable parametrization in terms of the recoil energy and the target parameters, namely:
\beq
\upsilon_{min}=\sqrt{\frac{A \,m_p \,E_R}{2 \, \mu^2_r}}
\eeq
 where $A m_p$ is the mass of the nucleus, $\mu_r $ is the reduced mass of the WIMP-nucleus system and $E_R$ is the energy transfer to the nucleus. 
\end{itemize}

With the above procedure  one obtains the quantity  $g(\upsilon_{min})$. For the M-B distribution in the local frame it is defined as follows:
\beq
g(\upsilon_{min},\upsilon_E(\alpha))=\frac{1}{\left (\sqrt{\pi}\upsilon_0 \right )^3}\int_{\upsilon_{min}}^{\upsilon_{max}}e^{-(\upsilon^2+2 \vbf . \vbf_E(\alpha)+\upsilon_E^2(\alpha))/\upsilon^2_0} \, \upsilon  \,  d\upsilon  \, d \Omega,\quad\upsilon_{max}=\upsilon_{esc}
\eeq 
$\vbf_E(\alpha)$ is the velocity of the Earth, including the velocity of the Sun  around the galaxy, $\vbf_E(\alpha)= \epsilon_0({\hat\upsilon}_s+\delta \left (\sin{\alpha}{\hat x}-\cos{\alpha}\cos{\gamma}{\hat y}+\cos{\alpha}\sin{\gamma} {\hat \upsilon}_s\right ))$ (see Eq. (\ref{Eq:vlocal})). The above upper cut off value in the M-B is usually put in by hand. Such a cut off comes in naturally, however, in the case of velocity distributions obtained from the halo WIMP mass density in the Eddington approach \cite{VEROW06}, which, in certain models, resemble a M-B distribution \cite{JDV09}. Its precise value is not, however, important for the results of the present paper.\\
Even though the differential rate is proportional \cite{spergel12} to $g(\upsilon_{min},\upsilon_E(\alpha))$, for the benefit of the experimentalists, we would like to make more explicit the dependence of the  differential rate on each of the variables entering the expression $g(\upsilon_{min},\upsilon_E(\alpha))$ and in particular to isolate the coefficient of $\cos{\alpha}$ term, which will provide the interesting modulation amplitude and make the time dependence explicit. This approach will be even more useful, when one integrates the differential rate to obtain the total event rate.\\
To this end, we will find it useful to expand $g(\upsilon_{min},\upsilon_E(\alpha))$ in powers of $\delta$, keeping terms up to linear in $\delta \approx 0.135$. We found it convenient to  express all velocities in units of the Sun's velocity $\upsilon_0$ to  obtain:
\beq
\upsilon_0 \, g(\upsilon_{min},\upsilon_E(\alpha))=\Psi_0(x)+\Psi_1(x)\cos{\alpha}, \quad x=\frac{\upsilon_{min}}{\upsilon_{0}}
\eeq
$\Psi_0(x)$ represents the quantity relevant for the average rate and $\Psi_1(x)$, which is proportional to $\delta$, represents the effects of modulation. \\
In the case of a M-B distribution these functions take the following form:
\beq
\Psi_0(x)=\frac{1}{2}
  \left [\mbox{erf}(1-x)+\mbox{erf}(x+1)+\mbox{erfc}(1-y_{\mbox{\tiny{esc}}})+\mbox{erfc}(y_{\mbox{\tiny{esc}}}+1)-2 \right ]
  \label{Eq:Psi0MB}
\eeq
\barr
\Psi_1(x)&=&\frac{1}{2} ~\delta 
   \left[\frac{ -\mbox{erf}(1-x)-\mbox{erf}(x+1)-\mbox{erfc}(1-y_{\mbox{\tiny{esc}}})-
   \mbox{erfc}(y_{\mbox{\tiny{esc}}}+1)}{2} \right . \nonumber\\
  && \left . +\frac{ e^{-(x-1)^2}}{\sqrt{\pi }}
   +\frac{
   e^{-(x+1)^2}}{\sqrt{\pi }}-\frac{ e^{-(y_{\mbox{\tiny{esc}}}-1)^2}}{\sqrt{\pi
   }}-\frac{ e^{-(y_{\mbox{\tiny{esc}}}+1)^2}}{\sqrt{\pi }}+1\right]
   \label{Eq:Psi1MB}
\earr
where erf$(x)$ and erfc$(x)$ are the error function and its complement, respectively, and $y_{esc}=\upsilon_{esc}/\upsilon_0\approx2.84$.

The differential event rate can be cast in the form:
\beq
\left .\frac{d R}{ d E_R}\right |_A=\left .\frac{dR_0}{dE_R}\right |_A+\left .\frac{dR_1}{dE_R}\right |_A \cos{\alpha}
\eeq
where the first term represents the time averaged (non modulated) differential event rate, while the second  gives the time dependent (modulated) one due to the motion of the Earth (see below). Furthermore for the coherent (here neutron coherent) in the case of a Dirac WIMP we get:
\barr
\left .\frac{d R_0}{ dE_R }\right |_A&=&\frac{\rho_{\chi}}{m_{\chi}}\,\frac{m_t}{A m_p}\, \left (\frac{\mu_r}{\mu_p} \right )^2\, \sqrt{<\upsilon^2>} \,\frac{1}{Q_0(A)} N^2 \sigma_N^{\mbox{\tiny{coh}}}\left .\left (\frac{d t}{du}\right ) \right |_{\mbox{ \tiny coh}} \nonumber\\
\left .\frac{d R_1}{ d E_R}\right |_A&=&\frac{\rho_{\chi}}{m_{\chi}}\,\frac{m_t}{A m_p} \,\left (\frac{\mu_r}{\mu_p} \right )^2\, \sqrt{<\upsilon^2>} \,\frac{1}{Q_0(A)} N^2 \sigma_N^{\mbox{\tiny{coh}}}\left .\left (\frac{d h}{du}\right )  \right |_{\mbox{\tiny coh}} 
\label{drdu}
\earr
with with $\mu_r$ ($\mu_p$) the WIMP-nucleus (nucleon) reduced mass and $A$ is the nuclear mass number. $ m_{\chi}$ is the WIMP mass, $\rho(\chi)$ is the WIMP density in our vicinity, assumed to be 0.3 GeV cm$^{-3}$,  and $m_t$ the mass of the target. 
Sometimes we will write the differential rate as:
\beq
\left .\frac{d R}{ dE_R }\right |_A=\frac{\rho_{\chi}}{m_{\chi}}\,\frac{m_t}{A m_p}\,  \left ( \frac{\mu_r}{\mu_p} \right )^2 \sqrt{<\upsilon^2>} \,\frac{1}{Q_0(A)} \,N^2 \,\sigma_N^{\mbox{\tiny{coh}}}\,\left .\left ( \frac{d t} {du}\right )\right |_{\mbox{ \tiny coh}}\,\left (1+ H(a \sqrt{u}) \cos{\alpha}\right )
\label{dhduH}
\eeq
In this formulation $H(a \sqrt{u}) $, the ratio of the modulated to the non modulated differential rate, gives the relative differential modulation amplitude, which is independent of the elementary cross section and the nuclear physics.

\begin{figure}
\begin{center}
\subfloat[]
{
\rotatebox{90}{\hspace{0.0cm} $F^2(E_R/Q0(A))\rightarrow$}
\includegraphics[width=0.4\textwidth]{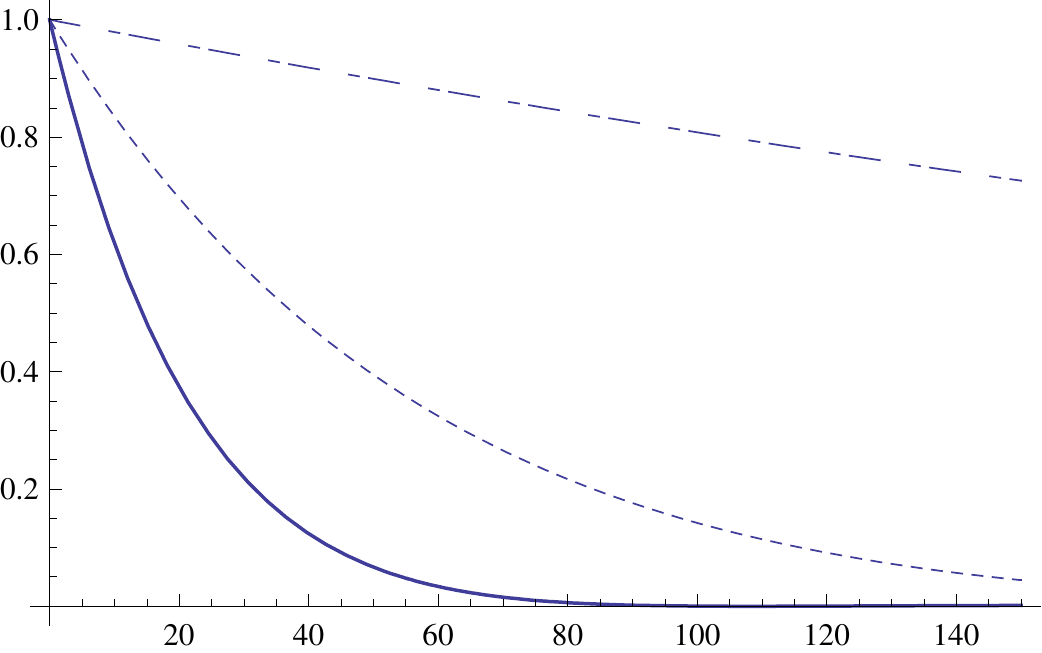}
}
\subfloat[]
{
\rotatebox{90}{\hspace{0.0cm} $F_{11}\rightarrow$}
\includegraphics[width=0.4\textwidth]{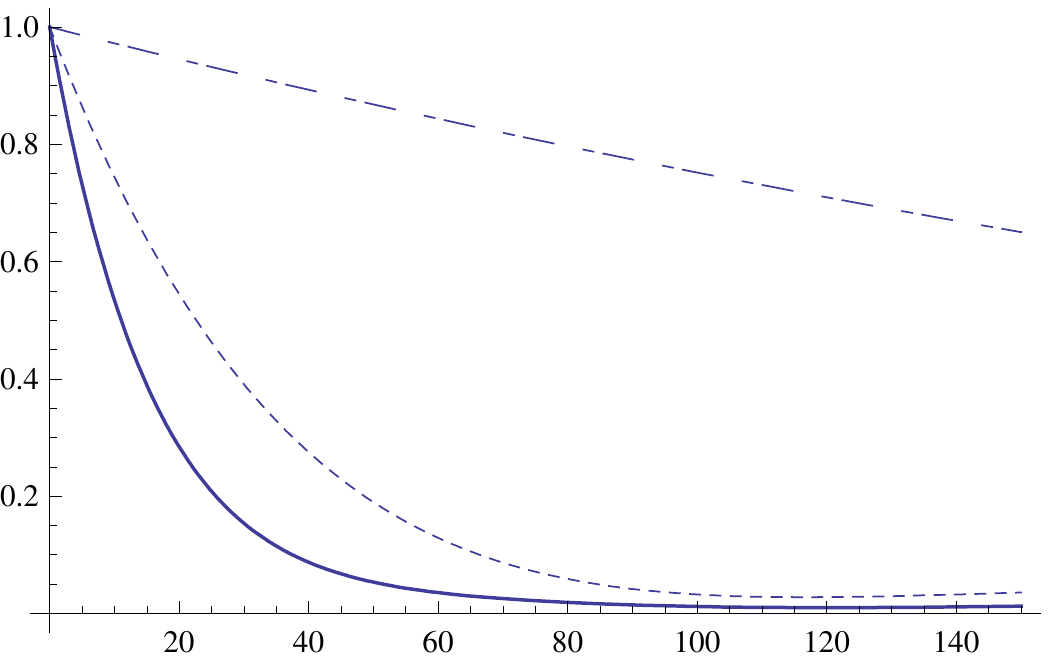}
}
\\
{\hspace{-2.0cm} $E_R \rightarrow$keV}
\caption{ The  square of the form factor entering the coherent scattering (a) and the spin response function in the isovector mode (b). With our normalization the other isospin modes are almost identical to the one shown. In both panels the solid, dotted and dashed curves correspond to the A=131, 73 and 19 systems respectively. The effect of the form factor for heavy targets becomes important even for an energy  transfer of  as low as 10 keV.
\label{fig:FFsq} }
\end{center}
\end{figure}

Furthermore one can show that
\beq
\left .\left (\frac{d t}{d u}\right )\,\right |_{\mbox{\tiny coh}}=\sqrt{\frac{2}{3}}\, a^2 \,F^2(u)  \, \Psi_0(a \sqrt{u}),\quad \left .\left (\frac{d h}{d u}\right )\right |_{\mbox{\tiny coh}}=\sqrt{\frac{2}{3}}\, a^2 \,F^2(u) \,\Psi_1(a \sqrt{u})
\eeq
The factor $\sqrt{2/3}$ is nothing but $\upsilon_0/\sqrt{\langle \upsilon ^2\rangle}$ since in Eq. (\ref{drdu}) $\sqrt{\langle \upsilon ^2\rangle}$ appears. In the above expressions  $a=(\sqrt{2} \mu_r b \upsilon_0)^{-1}$, $\upsilon_0$ the velocity of the sun around the center of the galaxy and $b$ the nuclear harmonic oscillator size parameter characterizing the nuclear wave function.  $ u$ is the energy transfer $E_R$ in dimensionless units given by
\begin{equation}
 u=\frac{E_R}{Q_0(A)}~~,~~Q_{0}(A)=[m_pAb^2]^{-1}=40A^{-4/3}\mbox{ MeV}
\label{defineu}
\end{equation}
and $F(u)$ is the nuclear form factor. Note that the parameter $a$ depends both on the WIMP mass, the target and the velocity distribution. Note also that for a given energy transfer $E_R$ the quantity $u$ depends on $A$. The square of the form factor is exhibited in Fig. \ref{fig:FFsq}.\\

For the axial current (spin induced) contribution, present for Dirac WIMPs and dominant for Majorana WIMPs one finds:
\barr
\left .\frac{d R_0}{ dE_R }\right |_A&=&\frac{\rho_{\chi}}{m_{\chi}}\,\frac{m_t}{A m_p}\, \left (\frac{\mu_r}{\mu_p} \right )^2\, \sqrt{<\upsilon^2>} \,\frac{1}{Q_0(A)} \frac{1}{3}\,\left ( \Omega_p-\Omega_n\right )^2\, \sigma_N^{\mbox{\tiny{spin}}}\left .\left (\frac{d t}{du}\right ) \right |_{\mbox{ \tiny spin}} \nonumber\\
\left .\frac{d R_1}{ d E_R}\right |_A&=&\frac{\rho_{\chi}}{m_{\chi}}\,\frac{m_t}{A m_p} \,\left (\frac{\mu_r}{\mu_p} \right )^2\, \sqrt{<\upsilon^2>} \,\frac{1}{Q_0(A)}\frac{1}{3}\,\left ( \Omega_p-\Omega_n\right )^2\, \sigma_N^{\mbox{\tiny{spin}}}\left .\left (\frac{d h}{du}\right )  \right |_{\mbox{\tiny spin}} 
\label{drdus}
\earr
or
\beq
\frac{d R}{ d E_R}|_A=\frac{\rho_{\chi}}{m_{\chi}}\frac{m_t}{A m_p}  \, \left ( \frac{\mu_r}{\mu_p} \right )^2 \sqrt{<\upsilon^2>}\, \frac{1}{Q_0(A)}\frac{1}{3}\,\left ( \Omega_p-\Omega_n\right )^2\,\sigma_N^{\mbox{\tiny{spin}}}\left .\left (\frac{d t} {du}\right )\right |_{\mbox{ \tiny spin}}\left (1+ H(a \sqrt{u}) \cos{\alpha}\right )
\label{sdhduH}
\eeq
with
\beq
\left . \left (\frac{d t}{d u}\right )\right |_{\mbox{\tiny spin}}=\sqrt{\frac{2}{3}} \, a^2  \, F_{11}(u)   \,  \Psi_0(a \sqrt{u}),\quad \left .\left (\frac{d h}{d u}\right )\right |_{\mbox{\tiny spin}}=\sqrt{\frac{2}{3}}  \, a^2  \, F_{11}(u)  \, \Psi_1(a \sqrt{u}),
\eeq
where $F_{11}$ is the spin response function (the square of the spin form factor). The behavior of the spin responce function $F_{11}$ for the isovector (isospin 1) channel is exhibited in Fig. \ref{fig:FFsq}. The other spin response functions $F_{01}$ and $F_{00}$ are, in our normalization, almost identical to the one shown. Note that, in the cases shown in Fig. \ref{fig:FFsq}, the spin response functions are not different from the square of the form factors entering the coherent mode. Thus, except for the scale of the event rates, the behavior of the coherent and the spin modes is almost identical. Anyway the quantity $H(a \sqrt{u})$ appearing in Eqs (\ref{dhduH})and (\ref{sdhduH}) is the same.

Sometimes, the target has multiple components, either different atomic elements as in DAMA and KIMS or in the xenon experiments where several isotopes may contribute. In such cases the rates are additive, and the dimensionless recoil variable $u$ must be normalized for each component:
\beq
\frac{dR}{dE_R}|_A\rightarrow \sum_i X_i\frac{dR}{dE_R}|_{A_i},\quad u\rightarrow u_i=\frac{E_R}{Q_0(A_i)}
\eeq
where $ X_i$ is the fraction of the component $A_i$ in the target.

Integrating the above differential rates we obtain the total rate including the time averaged rate  and the relative modulation amplitude $h$ for each mode  given by:
\beq
R_{\mbox{\tiny coh}}=\frac{\rho_{\chi}}{m_{\chi}} \, \frac{m_t}{A m_p}  \left ( \frac{\mu_r}{\mu_p} \right )^2  \sqrt{<\upsilon^2>} \, N^2 \, \sigma_N^{\mbox{\tiny{coh}}} \, t_{\mbox{\tiny coh}} \, \left (1+h_{\mbox{\tiny coh}} \cos{\alpha}\right ) ,
\label{Eq:Trates}
\eeq
\beq
R_{\mbox{\tiny spin}}=\frac{\rho_{\chi}}{m_{\chi}} \, \frac{m_t}{A m_p} \, \left ( \frac{\mu_r}{\mu_p} \right )^2  \sqrt{<\upsilon^2>} \, \frac{1}{3} \, \left ( \Omega_p-\Omega_n\right )^2 \, \sigma_N^{\mbox{\tiny{spin}}} \, t_{\mbox{\tiny spin}} \, \left (1+h_{\mbox{\tiny spin}} \cos{\alpha}\right ).
\label{Eq:Tratec}
\eeq
with
\beq
t=\int_{E_{th}/Q_0(A)}^{(y_{\mbox{\tiny esc}}/a)^2} \, \frac{dt}{du}\, du,\quad h=\frac{1}{t}\int_{E_{th}/Q_0(A)}^{(y_{\mbox{\tiny esc}}/a)^2} \, \frac{dh}{du} \, du
\label{Eq:thfac}
\eeq
for each mode (spin and coherent). $E_{th}(A)$ is the energy threshold imposed by the detector.

The spin ME can be  obtained in the context of a given nuclear model. Some such matrix elements of interest to the planned
experiments are given in table \ref{table.spin}, both in the isospin ($\Omega_0,\Omega_1 $) and the proton neutron ($\Omega_p,\Omega_n $) bases.
 The shown results
are obtained from DIVARI \cite{DIVA00}, Ressel {\it et al} (*) \cite{Ress},
 the Finish group (**) \cite{SUHONEN03}, (+)  \cite{Toivanen08}, .

\begin{figure}
\begin{center}
\subfloat[]
{
\includegraphics[width=0.35\textwidth]{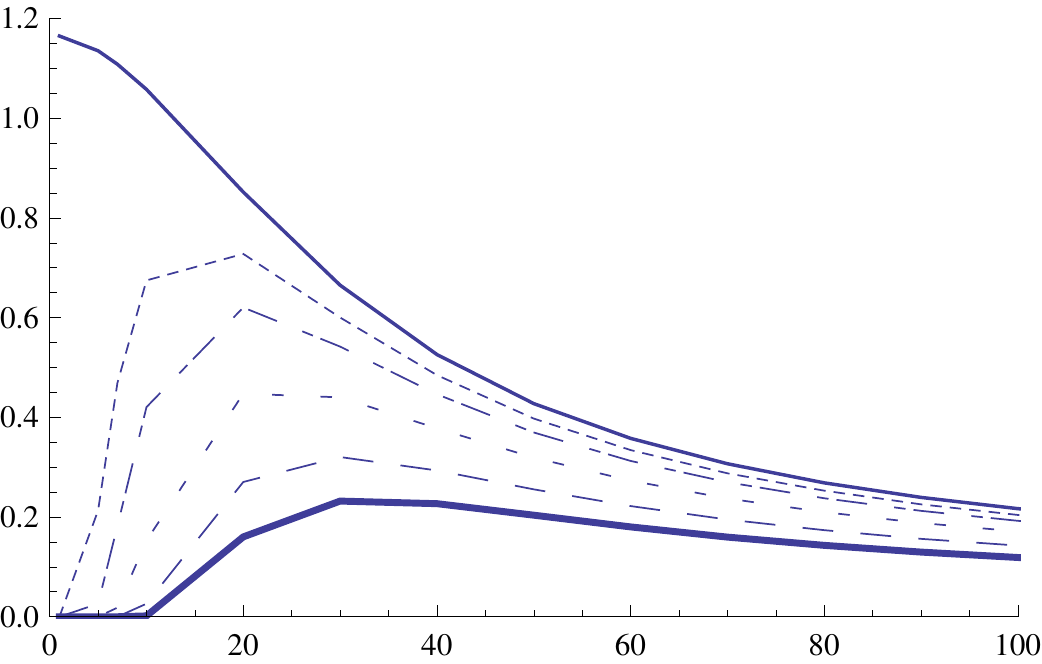}
}
\subfloat[]
{
\includegraphics[width=0.35\textwidth]{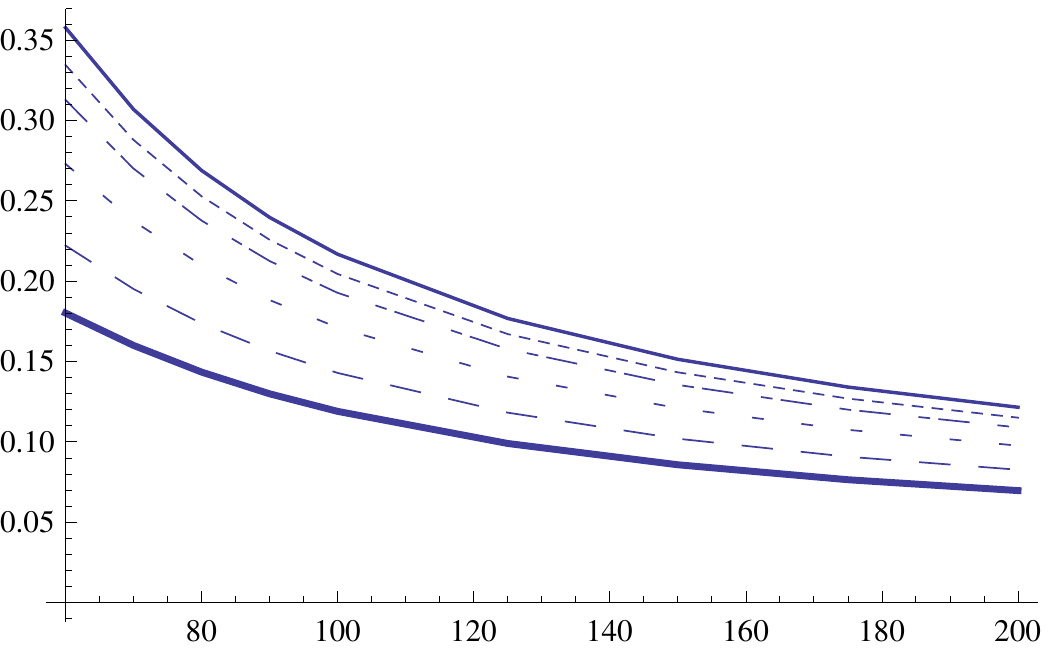}
}
\\
{\hspace{-2.0cm} $m_{\chi} \rightarrow$GeV}\\
\subfloat[]
{
\includegraphics[width=0.35\textwidth]{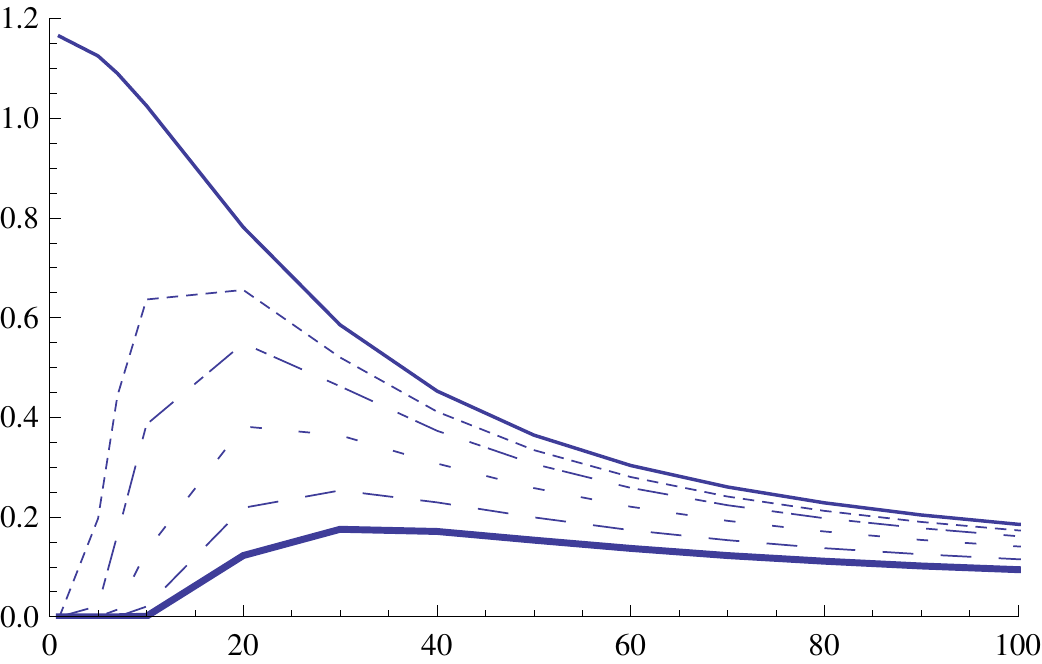}
}
\subfloat[]
{
\includegraphics[width=0.35\textwidth]{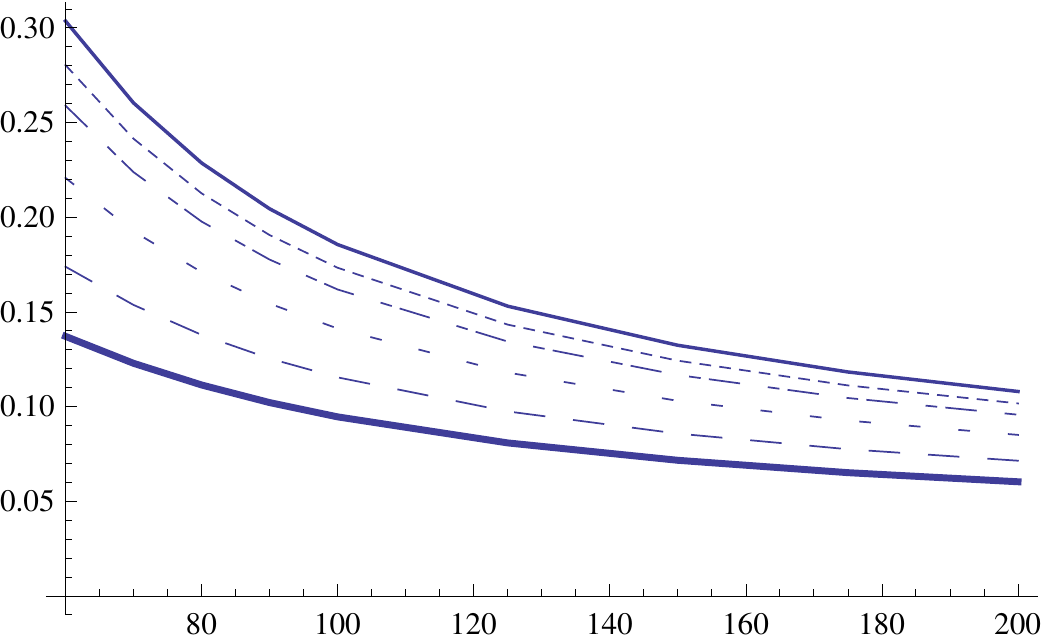}
}
\\
{\hspace{-2.0cm} $m_{\chi} \rightarrow$GeV}
\caption{ The parameters $t$ for the coherent (upper panels) and the spin (lower panels) in the case of the target A=131. On the right panels we exhibit these functions in the range of WIMP masses relevant to the present work, while on the left for comparison we show the low mass region. The curves correspond to  threshold energies 0,1,2,4,7,10 keV, with the threshold energy increasing from top to bottom. We see that even for a heavy nucleus, for which the form factor becomes important, there is very little difference between the coherent and the spin t-factors. 
 \label{fig:t131} }
\end{center}
\end{figure}

\begin{table}[t]
\caption{
 The static spin matrix elements for various nuclei. 
 For $^3$He see Moulin, Mayet and Santos
\cite{Santos}. For the other
light nuclei the calculations are from DIVARI \cite{DIVA00}.
 For  $^{73}$Ge and $^{127}$I the results presented  are from Ressel {\it et al}
\cite{Ress} (*), the Finnish group {\it et al} \cite {SUHONEN03} 
 (**) and
 the Finnish team (+) \cite{Toivanen08} and  Menendez (++) \cite{MeGazSCH12} .
 \label{table.spin} }
\begin{center}
\begin{tabular}{lrrrrrrrrrrrrr}
\hline\hline
 &   &  &  &  &   &  & & && &\\
 &$^3$ He& $^{19}$F & $^{29}$Si & $^{23}$Na  & $^{73}$Ge & $^{127}$I$^*$ & $ ^{127}$I$^{**}$ & $^{129}$Xe$^*$&$^{129}$Xe$^+$ & $^{129}$Xe$^{++}$ &$^{131}$Xe$^*$& $^{131}$Xe$^+$& $^{131}$Xe$^{++}$\\
\hline
    &   &  &  &  & & &  \\
$\Omega_{0}(0)$ &1.244     & 1.616   & 0.455  & 0.691  &1.075 & 1.815  &1.220  &1.341&-0.609 &1.174&-0.609& -0.325&-0.776\\
$\Omega_{1}(0)$&-1.527     & 1.675  & -0.461  & 0.588 &-1.003 & 1.105  &1.230  &-1.147&0.563&-1.105 &0.567&0.320&0.679\\
$\Omega_{p}(0)$ &-0.141    & 1.646  & -0.003  & 0.640  &0.036 &1.460   &1.225 &0.097 &-0.007&0.034 &-0.023&-0.002&-0.023\\
$\Omega_{n}(0)$ &1.386     & -0.030   & 0.459  & 0.051  &1.040 & 0.355  &-0.005 &1.244&0.946&1.140&0.567& -0.323&-0.702\\
\hline
\end{tabular}
\end{center}
\end{table}
Using the expressions for nucleon cross sections, (\ref{Eq:Trates}) and (\ref{Eq:Tratec}), we can obtain the total rates. These expressions contain the following parts: i) the parameters $t$ and $h$, which contain the effect of the velocity distribution and the nuclear form factors ii) the elementary nucleon cross sections iii) the nuclear physics input (nuclear spin spin ME, see  table \ref{table.spin}). \\ 
Let us briefly discuss the parameters $t$. These parameters depend on the reduced mass, velocity  distribution and, for a heavy nucleus, on  the nuclear form factors. In spite of this there is very little difference between the coherent and the spin contribution (see Fig. \ref{fig:t131}). The same thing applies in the case of the $h$ factor. The target dependence will be inferred from the total rates computed below.

\section{Some results for Majorana WIMPs}
 \label{sec:rates}
Before proceeding further we note that if the spin 3/2 particle were of the Dirac variety one would have a contribution arising from the time component of the neutral hadronic  current. This essentially zero for the proton component, but it would have lead to a coherent contribution on the neutron. In this case the rate would be very large for heavy targets, proportional to $N^2$, which is excluded by the current data. We will, therefore, concentrate on the Majorana case, for which the coherent component will be proportional to $\beta^2$, $\beta=\upsilon/c$, and is therefore negligible in comparison to the spin-dependent component. 

 We have seen that in our model the case of Dirac WIMPs is excluded. So we will concentrate on the Majorana spin 3/2 case.
 \subsection{The differential event rates} 
\begin{figure}
\begin{center}
\subfloat[]
{
\rotatebox{90}{\hspace{0.0cm} $\left .\left (\frac{dR_0}{dE_R}\right )\right|_A \rightarrow$(kg-y)/ keV}
\includegraphics[height=.25\textwidth]{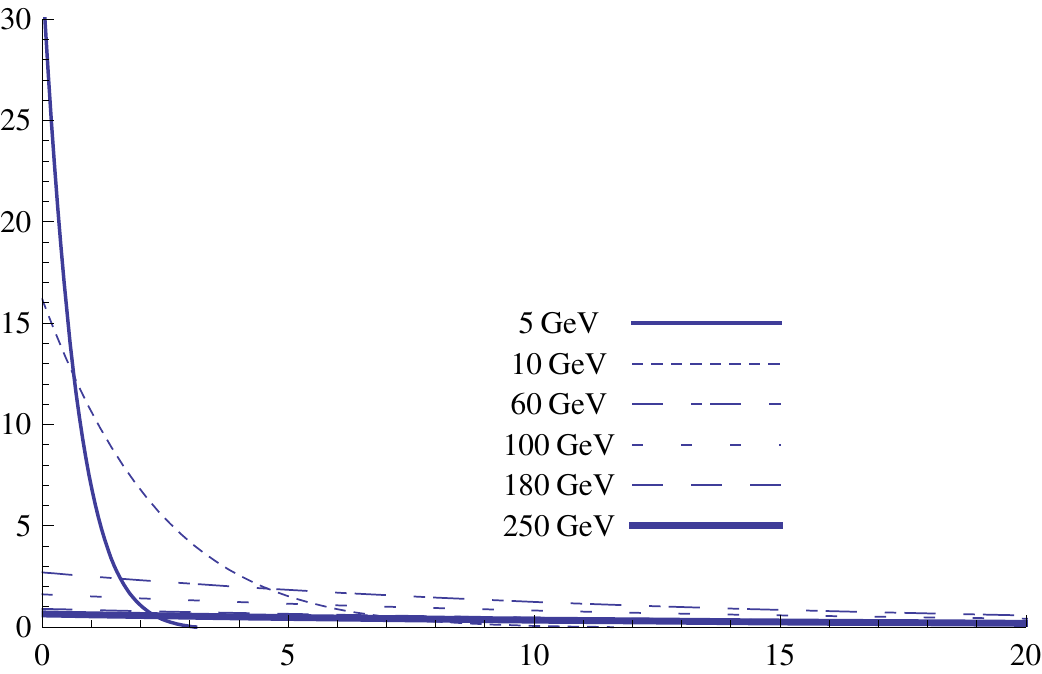}
}
\subfloat[]
{
\rotatebox{90}{\hspace{0.0cm} $\left .\left (\frac{dR_0}{dE_R}\right )\right|_A\rightarrow$(kg-y)/ keV)}
\includegraphics[height=.25\textwidth]{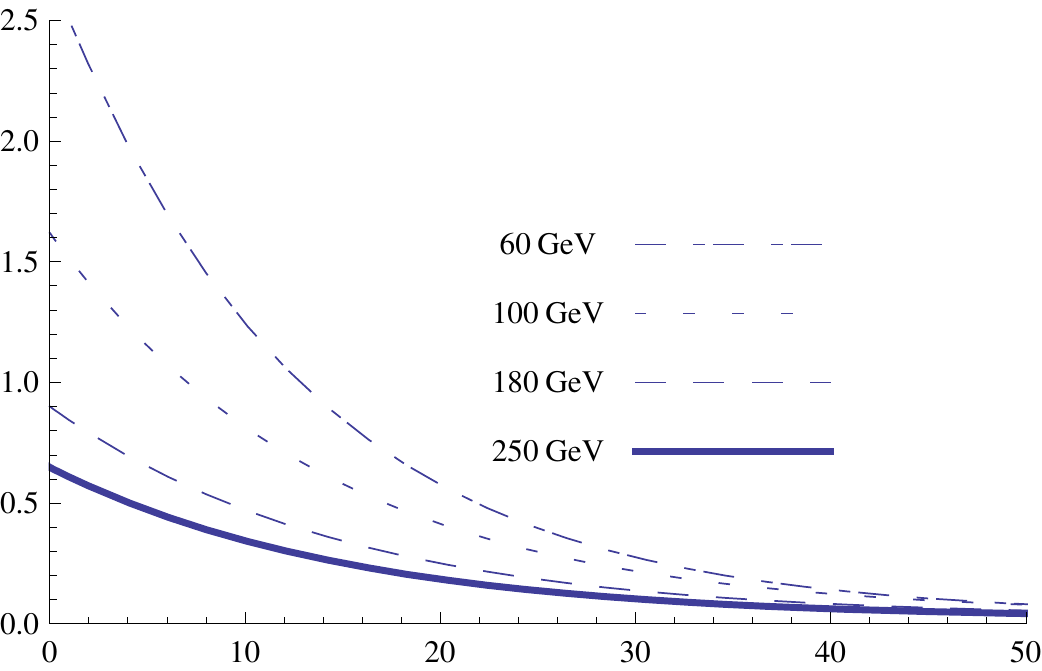}
}
\\
{\hspace{-2.0cm} $E_R\rightarrow$keV}\\

\subfloat[]
{
\rotatebox{90}{\hspace{0.0cm} $\left .\left (\frac{dR_1}{dE_R}\right )\right|_A \rightarrow$(kg-y)/ keV}
\includegraphics[height=.25\textwidth]{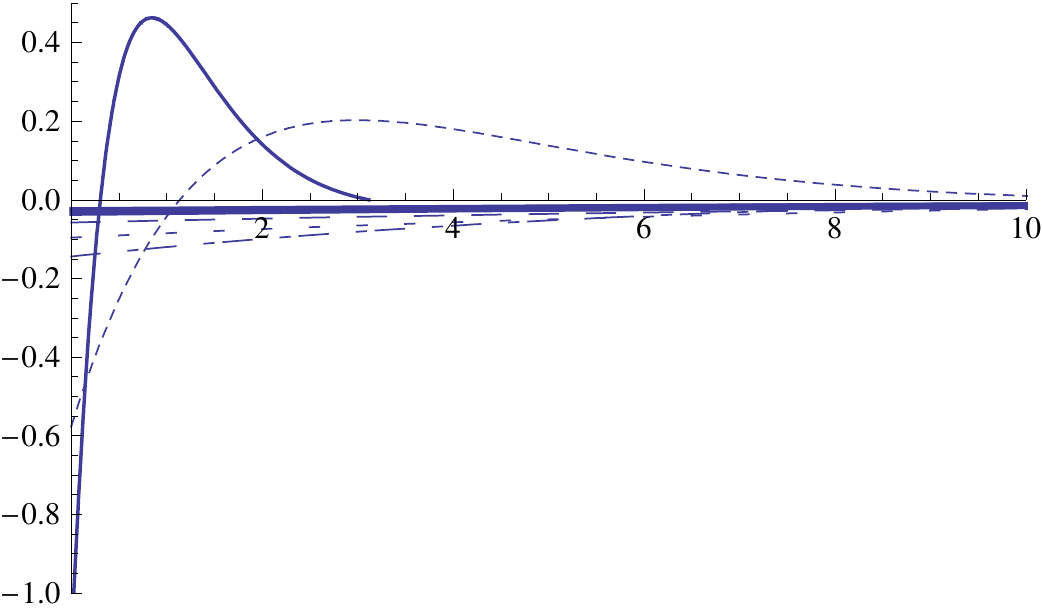}
}
\subfloat[]
{
\rotatebox{90}{\hspace{0.0cm} $\left .\left (\frac{dR_1}{dE_R}\right )\right|_A\rightarrow$(kg-y)/ keV)}
\includegraphics[height=.25\textwidth]{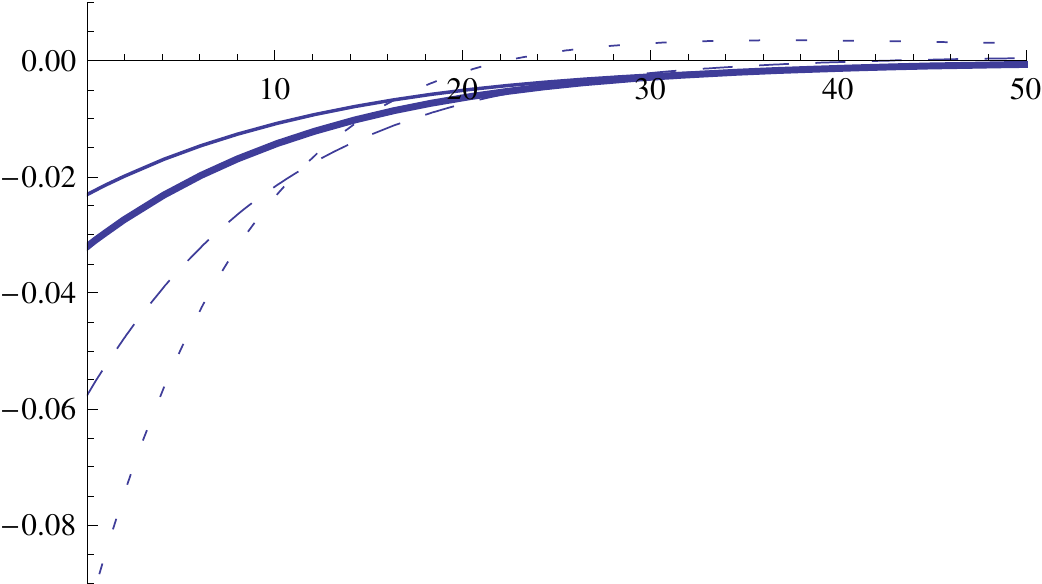}
}
\\
{\hspace{-2.0cm} $E_R\rightarrow$keV}\\
\caption{ On the top we show the time average differential rate  $\left .\frac{dR_0}{dE_R}\right|_A$,  and at the bottom panels we show the differential modulated amplitude $\left .\frac{dR_1}{dE_R}\right|_A$ as a function of the recoil energy $E_R$  in keV. These results correspond to the spin mode in the case of $^{131}$Xe for a Majorana WIMP. On the left panels  we exhibit curves for WIMP masses (5, 10,  60, 100, 180, 250) GeV, while on the right we exhibit WIMP masses relevant to our model, i.e. 60, 80, 180, 250 GeV. The relevant WIMP mass for each curve is explained on the top panels.
 \label{fig:dRhdQS_131}}
\end{center}
\end{figure}

\begin{figure}
\begin{center}
\subfloat[]
{
\rotatebox{90}{\hspace{0.0cm} $\left .\left (\frac{dR_0}{dE_R}\right )\right|_A \rightarrow$(kg-y)/ keV}
\includegraphics[height=.25\textwidth]{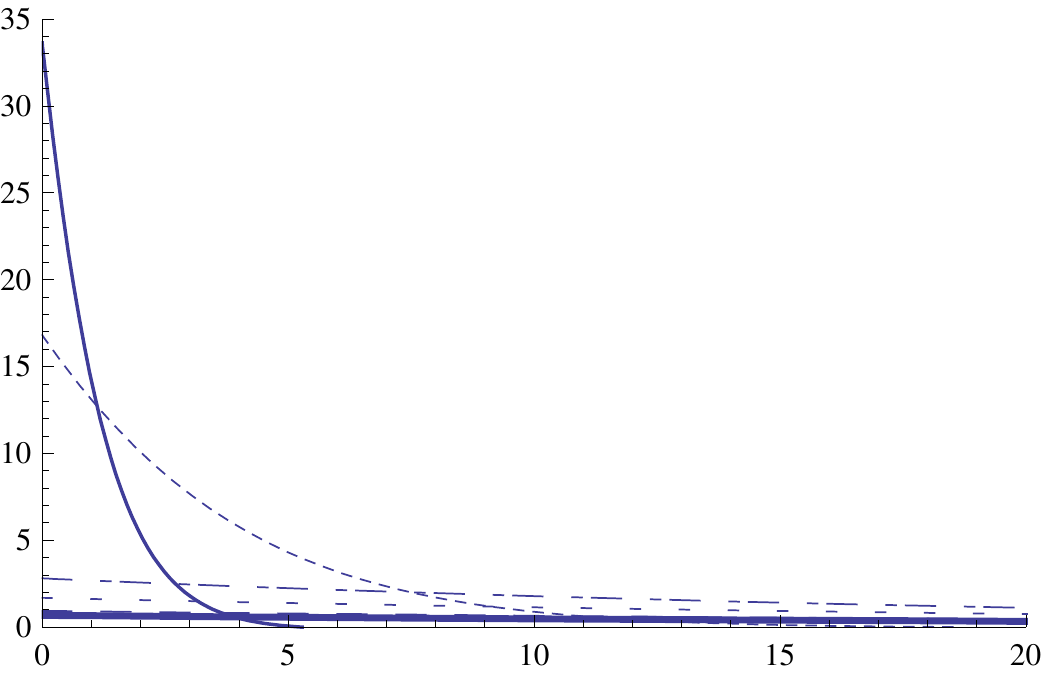}
}
\subfloat[]
{
\rotatebox{90}{\hspace{0.0cm} $\left .\left (\frac{dR_0}{dE_R}\right )\right|_A\rightarrow$(kg-y)/ keV)}
\includegraphics[height=.25\textwidth]{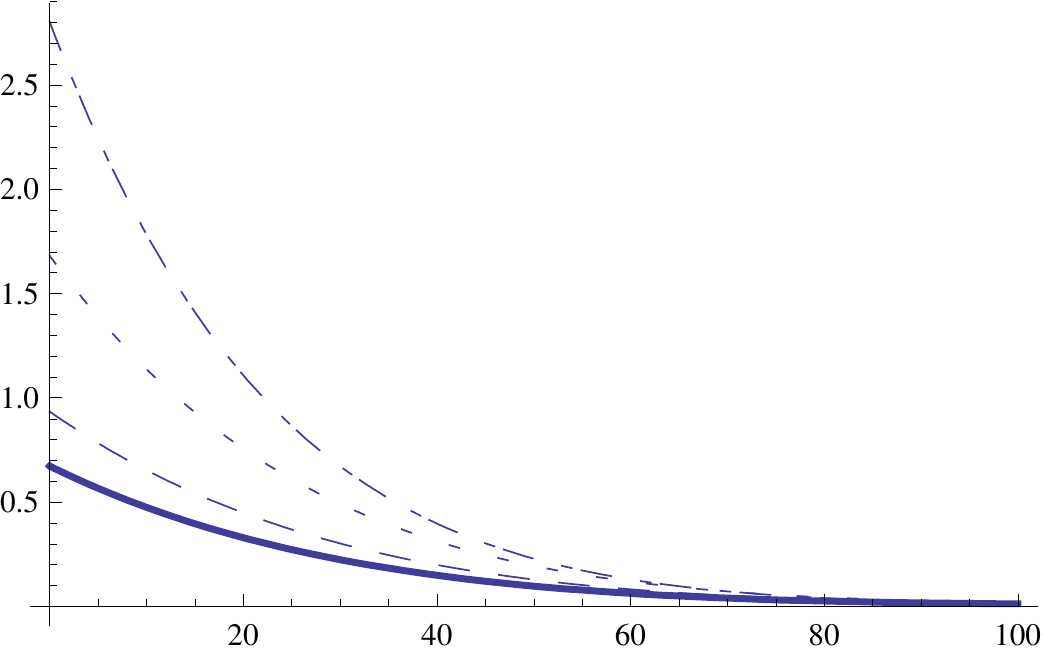}
}
\\
{\hspace{-2.0cm} $E_R\rightarrow$keV}\\

\subfloat[]
{
\rotatebox{90}{\hspace{0.0cm} $\left .\left (\frac{dR_1}{dE_R}\right )\right|_A \rightarrow$(kg-y)/ keV}
\includegraphics[height=.25\textwidth]{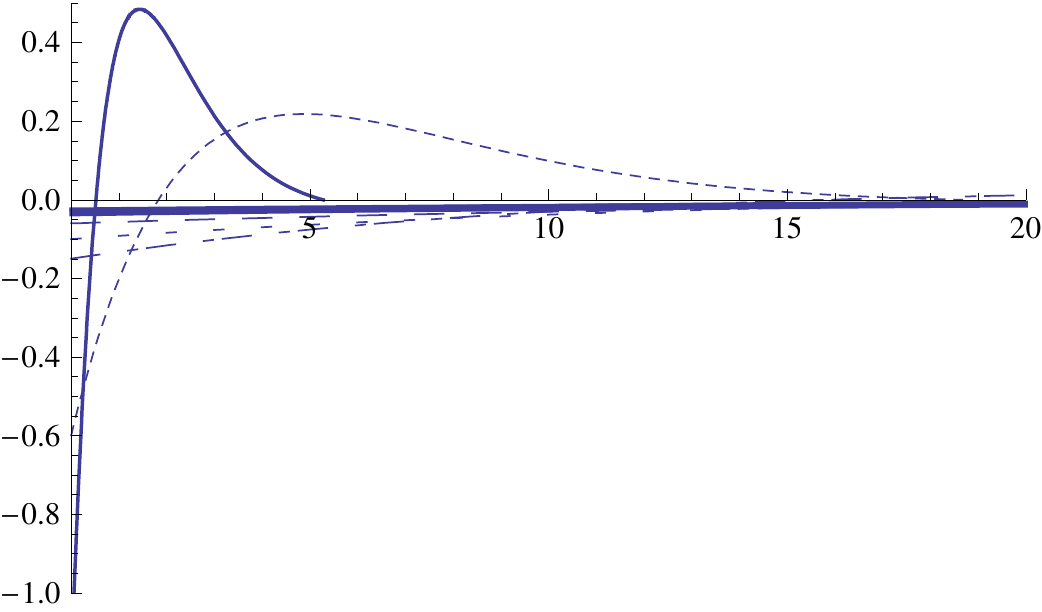}
}
\subfloat[]
{
\rotatebox{90}{\hspace{0.0cm} $\left .\left (\frac{dR_1}{dE_R}\right )\right|_A \rightarrow$(kg-y)/ keV}
\includegraphics[height=.25\textwidth]{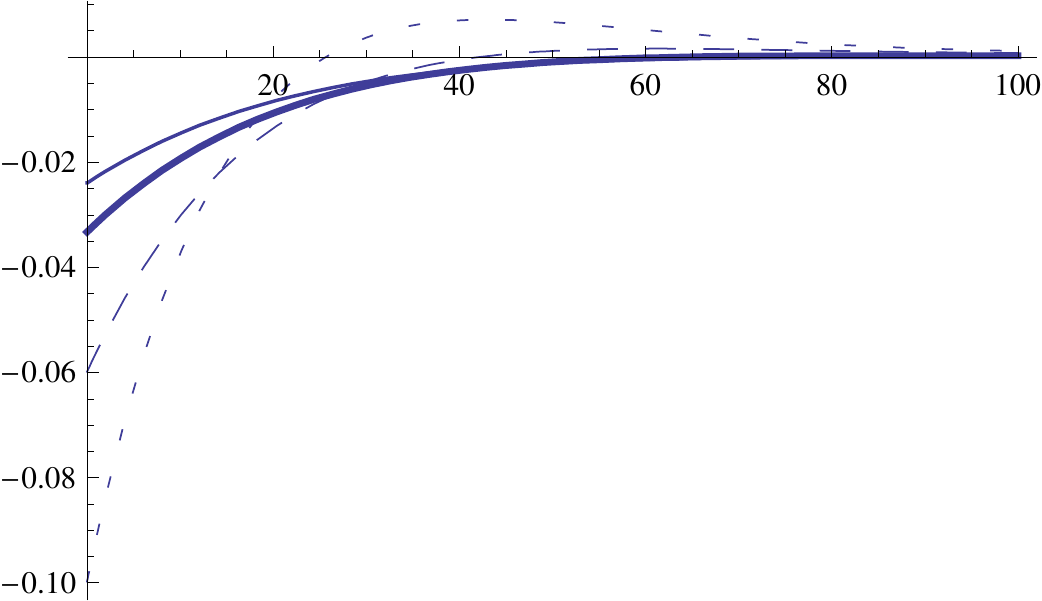}
}
\\
{\hspace{-2.0cm} $E_R\rightarrow$keV}\\
\caption{ The same as in Fig. \ref{fig:dRhdQS_131} for the A=73 target.
\label{fig:dRhdQS_73}}
\end{center}
\end{figure}

\begin{figure}
\begin{center}
\subfloat[]
{
\rotatebox{90}{\hspace{0.0cm} $\left .\left (\frac{dR_0}{dE_R}\right )\right|_A \rightarrow$(kg-y)/ keV}
\includegraphics[height=.25\textwidth]{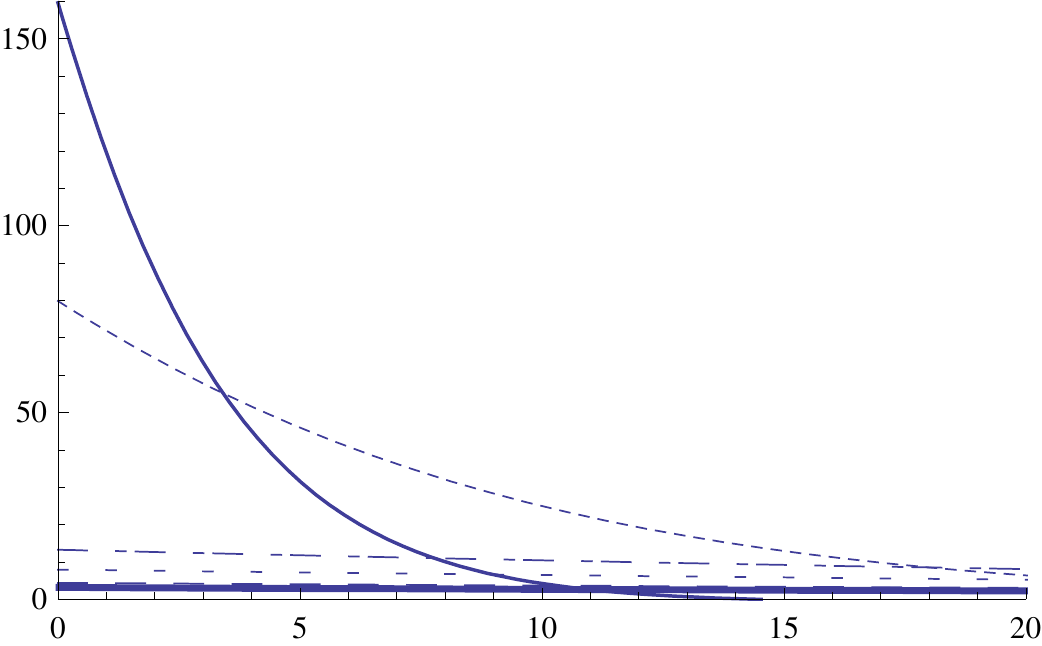}
}
\subfloat[]
{
\rotatebox{90}{\hspace{0.0cm} $\left .\left (\frac{dR_0}{dE_R}\right )\right|_A\rightarrow$(kg-y)/ keV)}
\includegraphics[height=.25\textwidth]{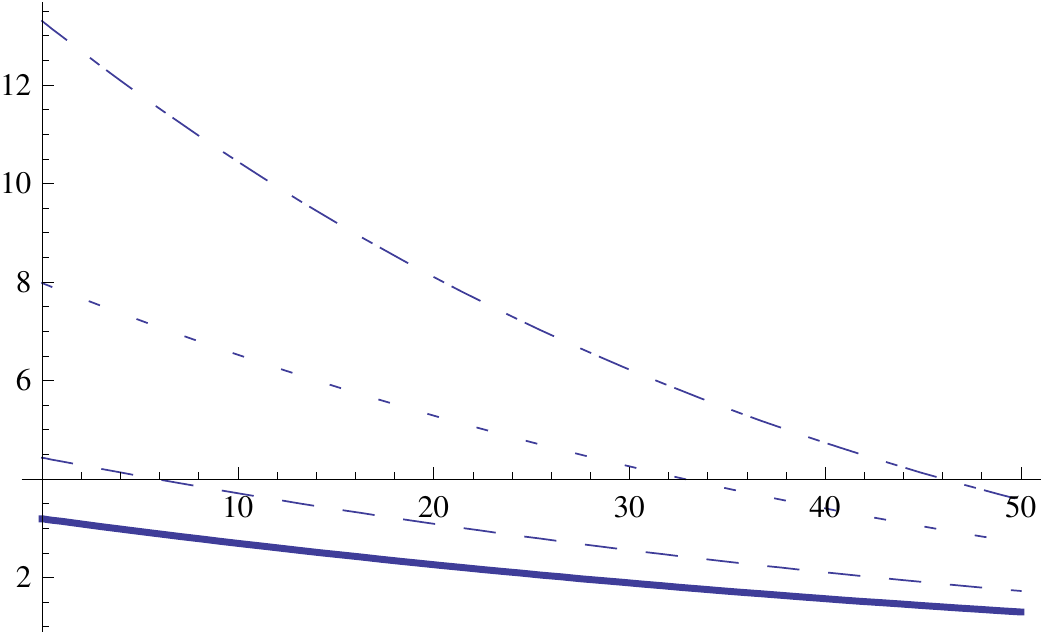}
}
\\
{\hspace{-2.0cm} $E_R\rightarrow$keV}\\

\subfloat[]
{
\rotatebox{90}{\hspace{0.0cm} $\left .\left (\frac{dR_1}{dE_R}\right )\right|_A \rightarrow$(kg-y)/ keV}
\includegraphics[height=.25\textwidth]{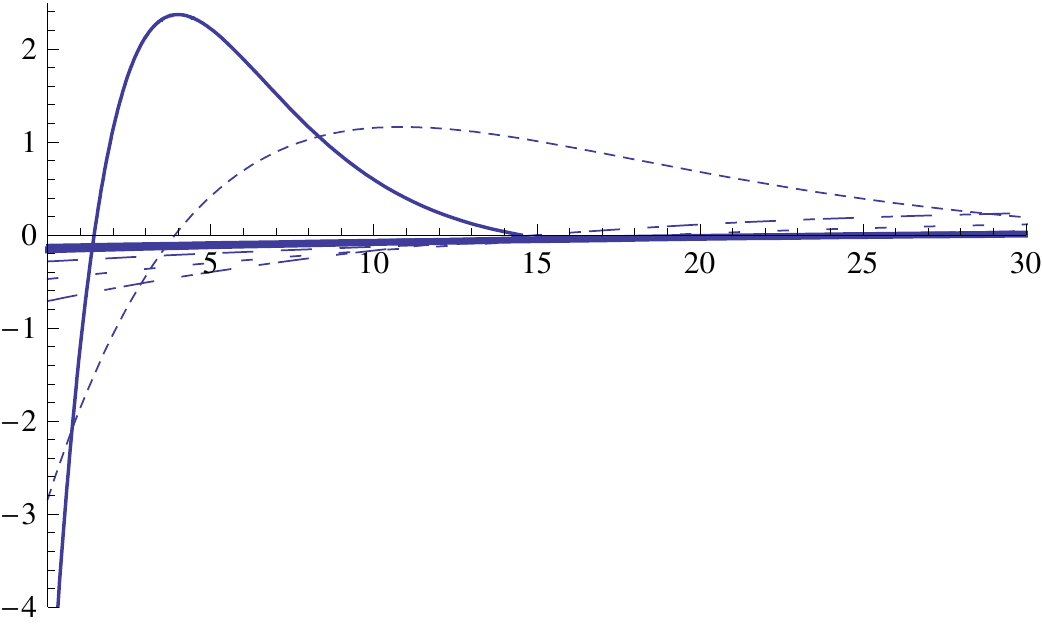}
}
\subfloat[]
{
\rotatebox{90}{\hspace{0.0cm} $\left .\left (\frac{dR_1}{dE_R}\right )\right|_A \rightarrow$(kg-y)/ keV}
\includegraphics[height=.25\textwidth]{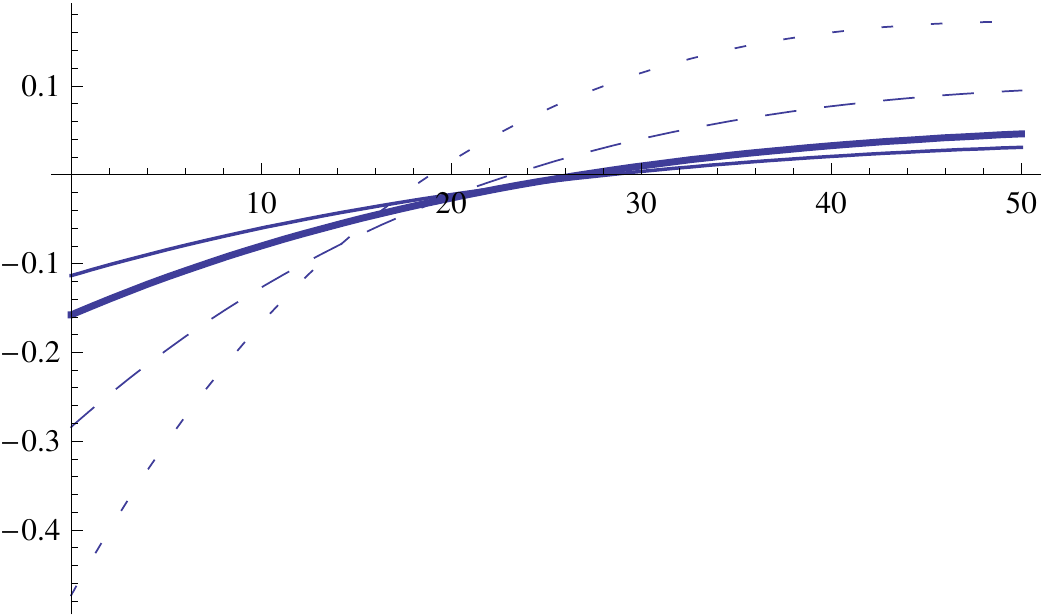}
}
\\
{\hspace{-2.0cm} $E_R\rightarrow$keV}\\
\caption{ The same as in Fig. \ref{fig:dRhdQS_131} for the A=19 target.
\label{fig:dRhdQS_19}}
\end{center}
\end{figure}

The differential event rates, perhaps the most  interesting from an experimental point of view,  depend on the WIMP mass.  So we can only present them for some select masses. Considerations based on the relic abundance of the WIMP in this model lead to the conclusion that it has a  mass between 80 and 200 GeV. This the WIMP mass range of interest to us.  For illustration purposes, however, we have decided to present some  results also for  lighter WIMPs. Our results for the  differential rates are  exhibited    in Figs \ref{fig:dRhdQS_131}- \ref{fig:dRhdQS_19}..
 \subsection{The total event rates}
\begin{figure}
\begin{center}
\subfloat[]
{
\includegraphics[width=0.4\textwidth]{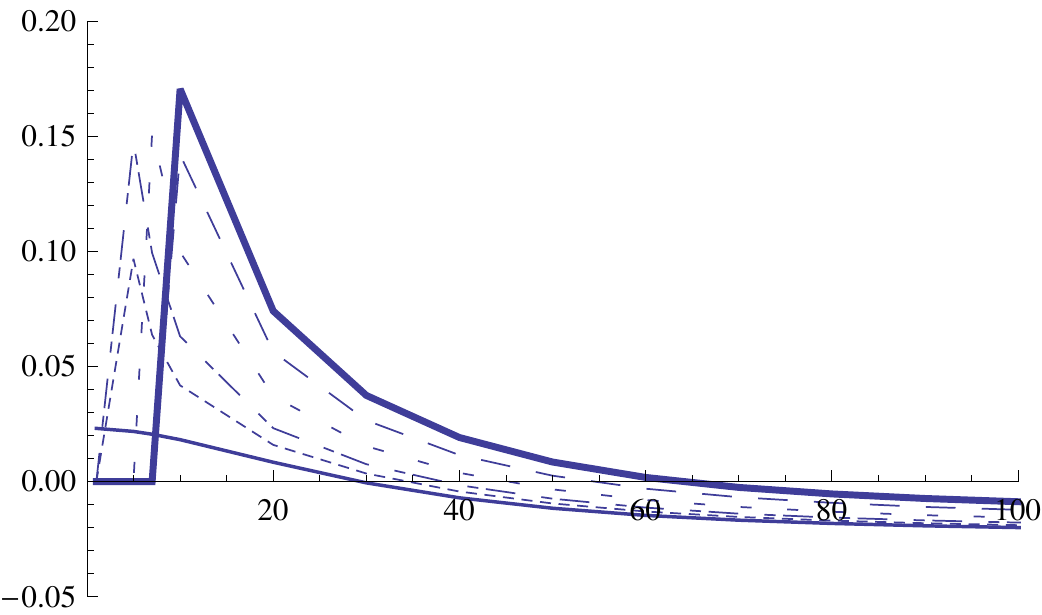}
}
\subfloat[]
{
\includegraphics[width=0.4\textwidth]{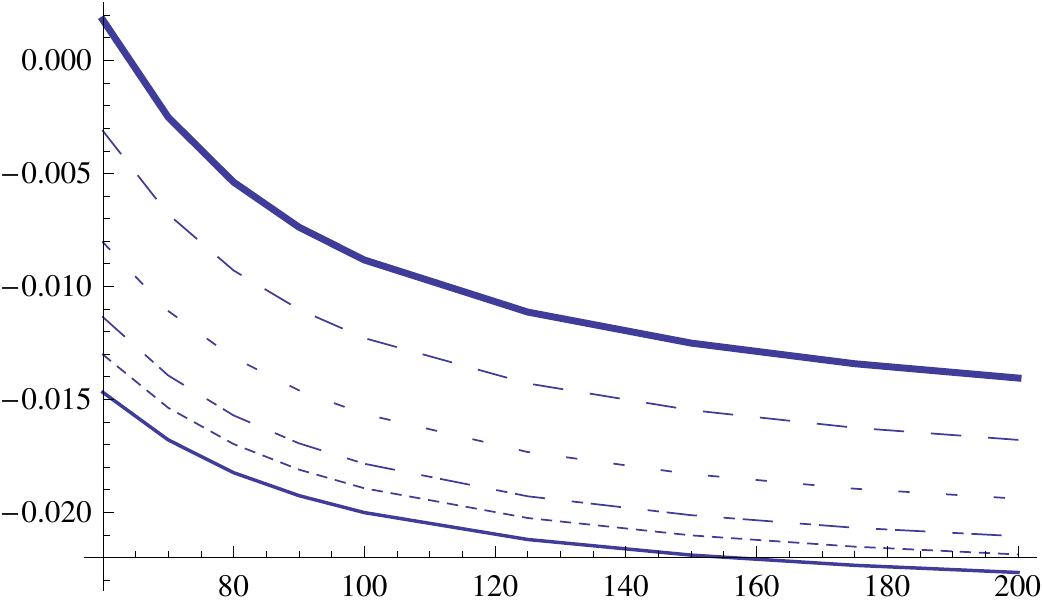}
}
\\
{\hspace{-2.0cm} $m_{\chi} \rightarrow$GeV}
\caption{ We show the relative modulation amplitude in the case of the spin for the target $^{131}$Xe. 
 Note the change in sign  for heavy WIMPs, meaning that for negative sign the maximum occurs six months later than naively expected.
As in  Fig. \ref{fig:t131}, the curves correspond to  threshold energies 0,1,2,4,7,10 keV, ordered generally from top to bottom.
 \label{Fig:hspin131} }
\end{center}
\end{figure}

\begin{figure}
\begin{center}
\subfloat[]
{
\includegraphics[width=0.4\textwidth]{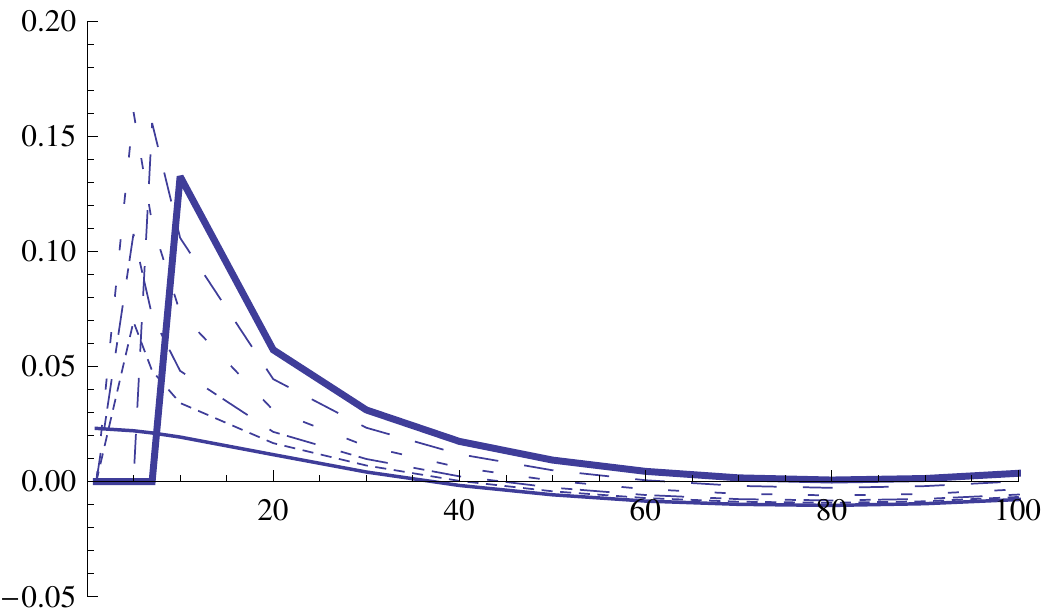}
}
\subfloat[]
{
\includegraphics[width=0.4\textwidth]{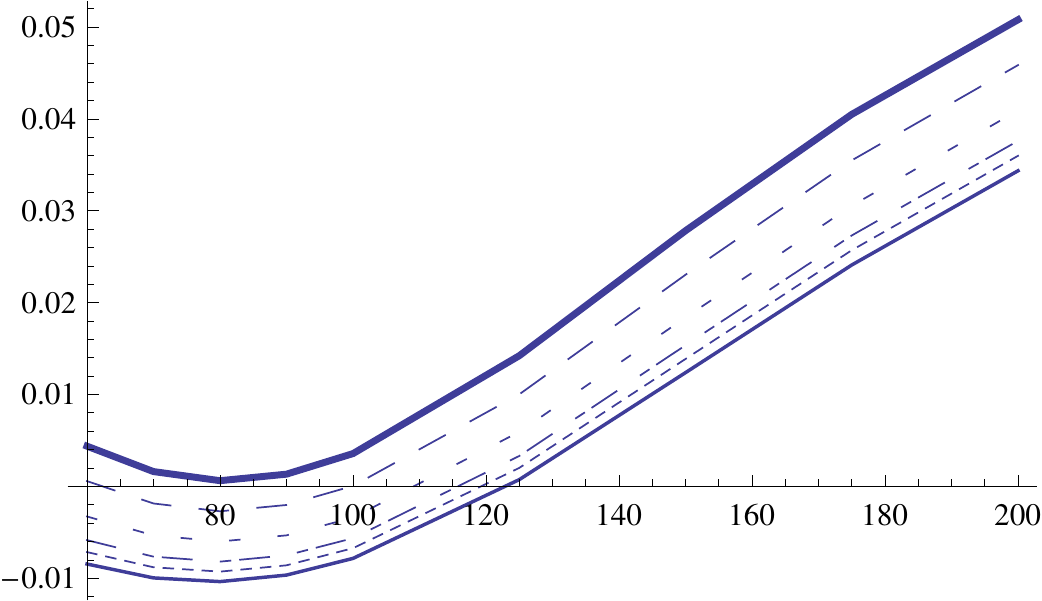}
}
\\
{\hspace{-2.0cm} $m_{\chi} \rightarrow$GeV}
\caption{ The same as in Fig. \ref{Fig:hspin131} for the target $^{73}$Ge. 
 \label{Fig:hspin73} }
\end{center}
\end{figure}

\begin{figure}
\begin{center}
\subfloat[]
{
\includegraphics[width=0.4\textwidth]{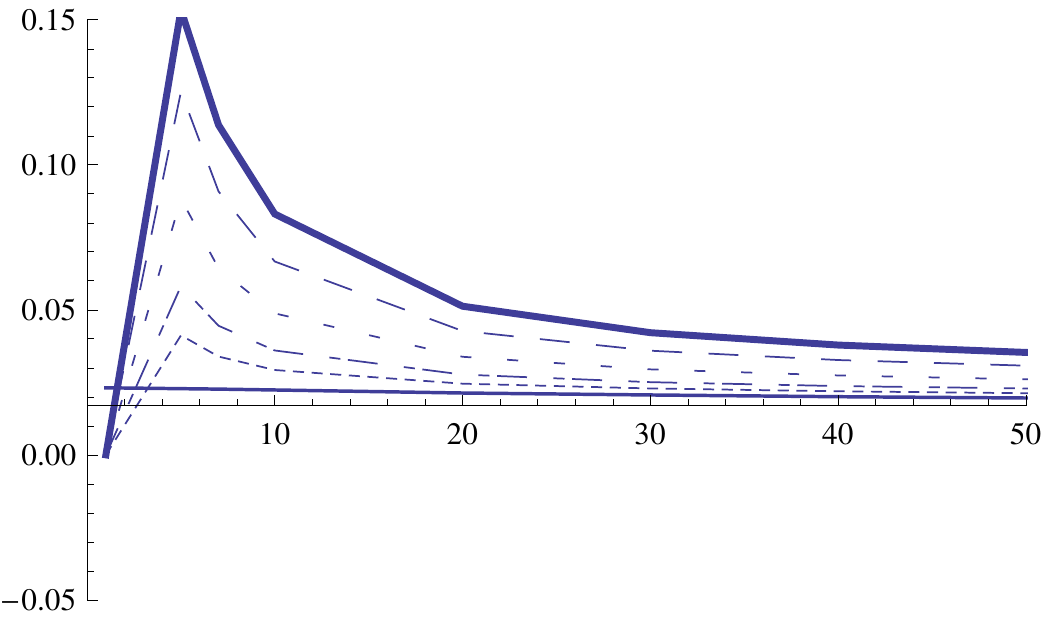}
}
\subfloat[]
{
\includegraphics[width=0.4\textwidth]{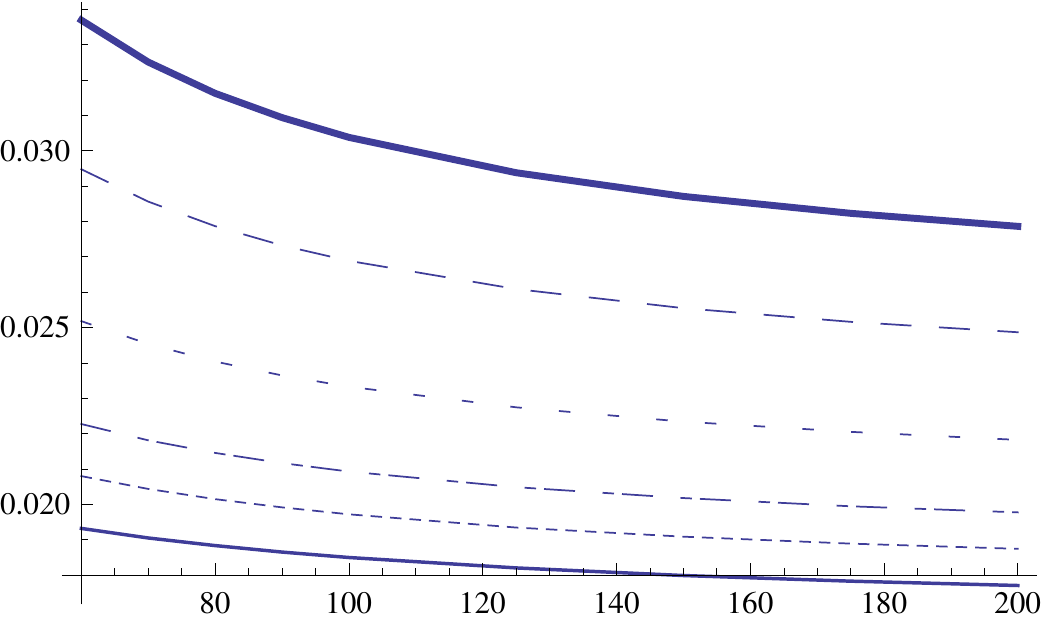}
}
\\
{\hspace{-2.0cm} $m_{\chi} \rightarrow$GeV}
\caption{ The same as in Fig. \ref{Fig:hspin131} for the target $^{19}$F. 
 \label{Fig:hspin19} }
\end{center}
\end{figure}

  We recall that the parameters $h_{\mbox{\tiny{coh}}}$ and $h_{\mbox{\tiny{spin}}}$ provide the ratio of the time varying amplitude divided by the time averaged rate. The relative modulated rate is given by $h_{coh}\cos{\alpha}$ and $h_{spin}\cos{\alpha}$, with $\alpha$ the phase of the Earth. In the case of the spin we will present these functions in Figs \ref{Fig:hspin131}-\ref{Fig:hspin19}.


\begin{figure}
\begin{center}
\subfloat[]
{
\rotatebox{90}{\hspace{0.0cm}$R_{\mbox{\tiny spin}}\rightarrow$ events/kg-y}
\includegraphics[width=0.4\textwidth]{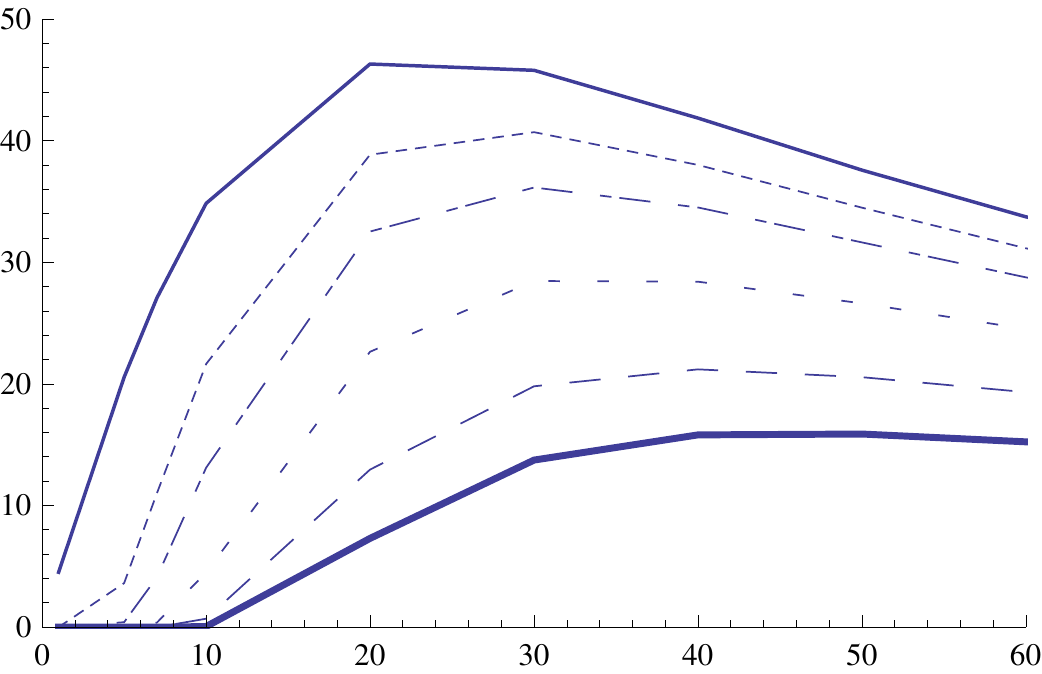}
}
\subfloat[]
{
\rotatebox{90}{\hspace{0.0cm} $R_{\mbox{\tiny spin}}\rightarrow$events/kg-y}
\includegraphics[width=0.4\textwidth]{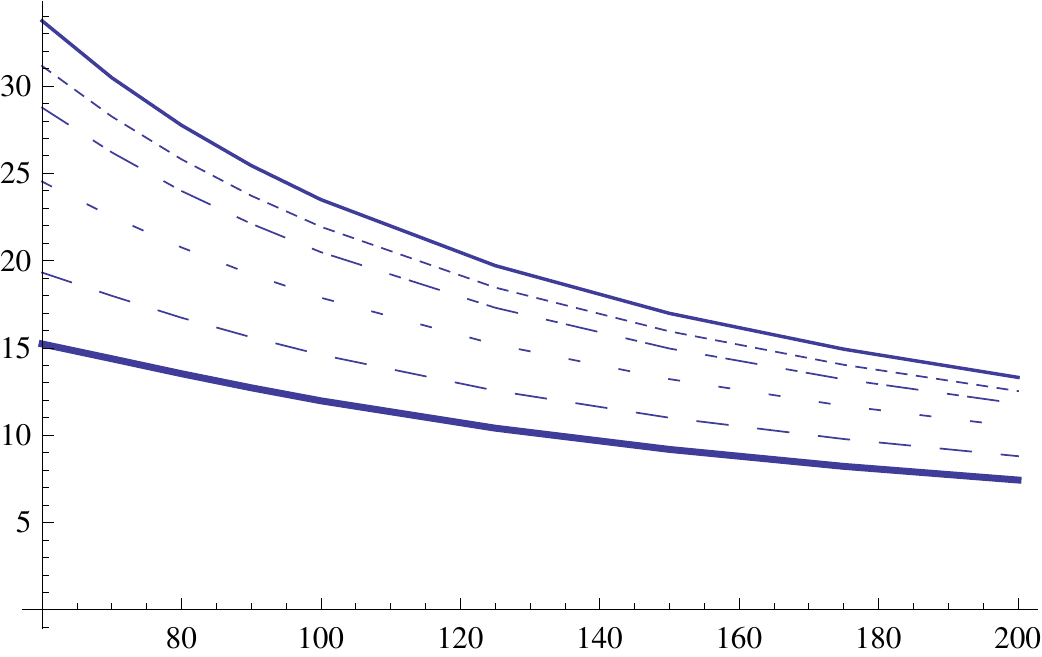}
}
\\
{\hspace{-2.0cm} $m_{\chi} \rightarrow$GeV}
\caption{ The  predicted time averaged total rate $R_{\mbox{\tiny spin}}$ for $^{131}$Xe  restricted to the small WIMP mass regime  (a) and its restriction to the   WIMP mass range  relevant to our model (b). Otherwise the notation is the same as that of Fig. \ref{fig:t131}.
 \label{fig:plotRs131}} 
\end{center}
\end{figure}

\begin{figure}
\begin{center}
\subfloat[]
{
\rotatebox{90}{\hspace{0.0cm}$R_{\mbox{\tiny spin}}\rightarrow$ events/kg-y}
\includegraphics[width=0.4\textwidth]{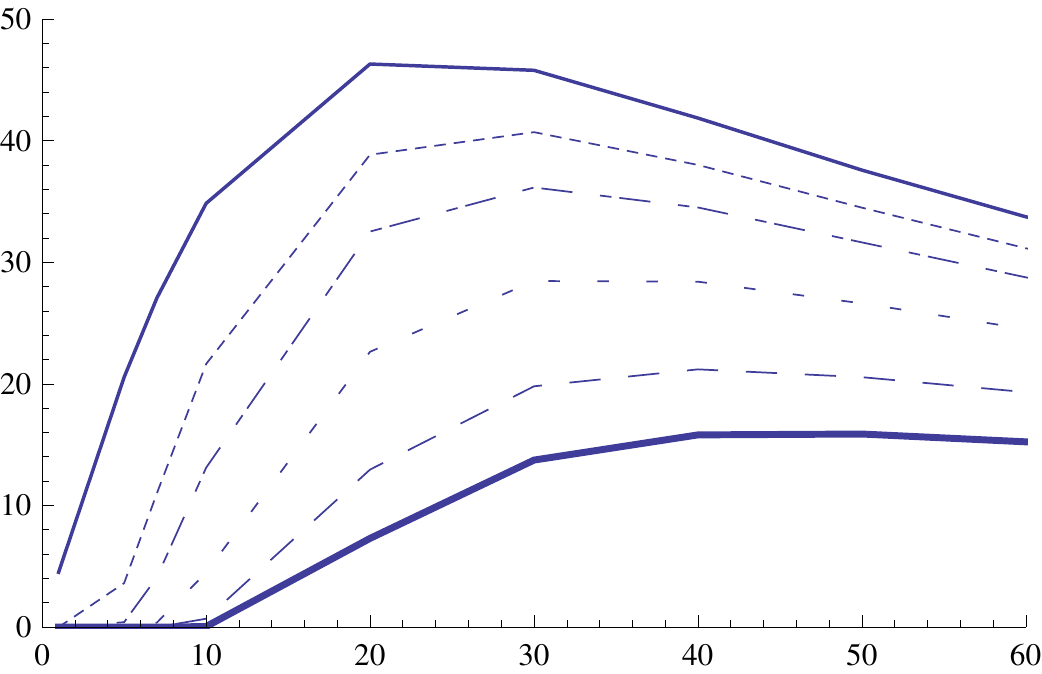}
}
\subfloat[]
{
\rotatebox{90}{\hspace{0.0cm} $R_{\mbox{\tiny spin}}\rightarrow$events/kg-y}
\includegraphics[width=0.4\textwidth]{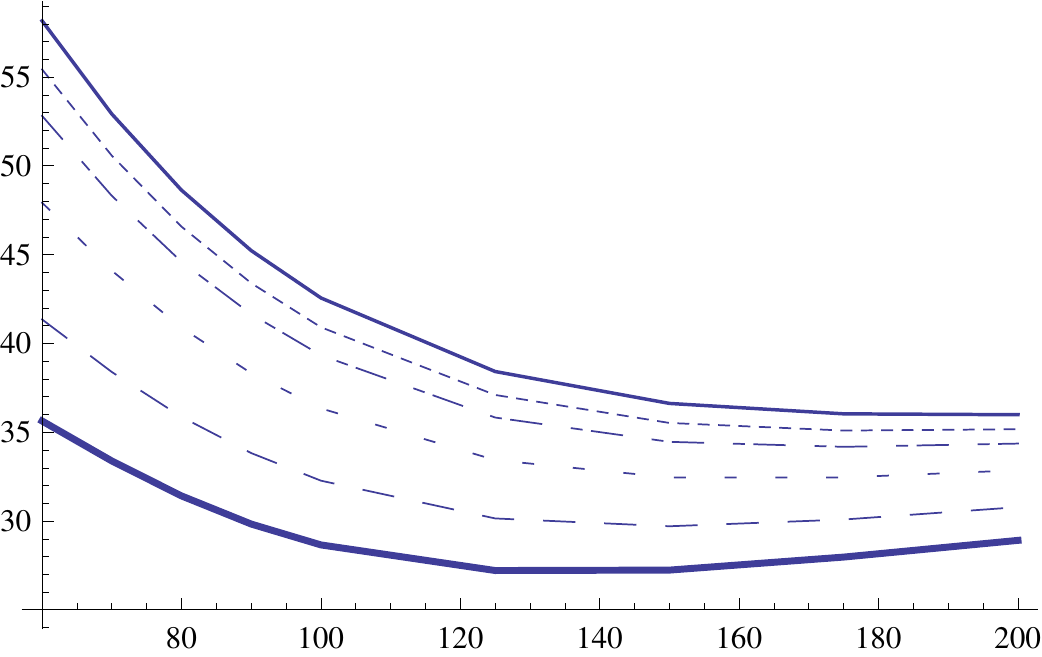}
}
\\
{\hspace{-2.0cm} $m_{\chi} \rightarrow$GeV}
\caption{ The  same as in Fig.  \ref{fig:plotRs131}for $^{73}$Ge.
 \label{fig:plotRs73}} 
\end{center}
\end{figure}

\begin{figure}
\begin{center}
\subfloat[]
{
\rotatebox{90}{\hspace{0.0cm}$R_{\mbox{\tiny spin}}\rightarrow$ events/kg-y}
\includegraphics[width=0.4\textwidth]{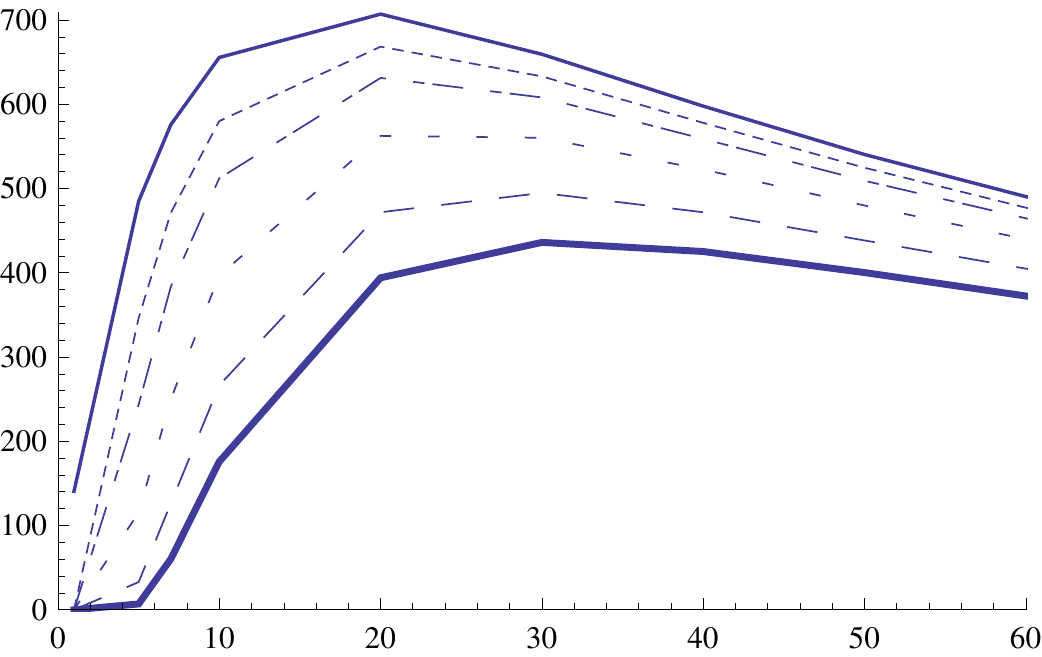}
}
\subfloat[]
{
\rotatebox{90}{\hspace{0.0cm} $R_{\mbox{\tiny spin}}\rightarrow$events/kg-y}
\includegraphics[width=0.4\textwidth]{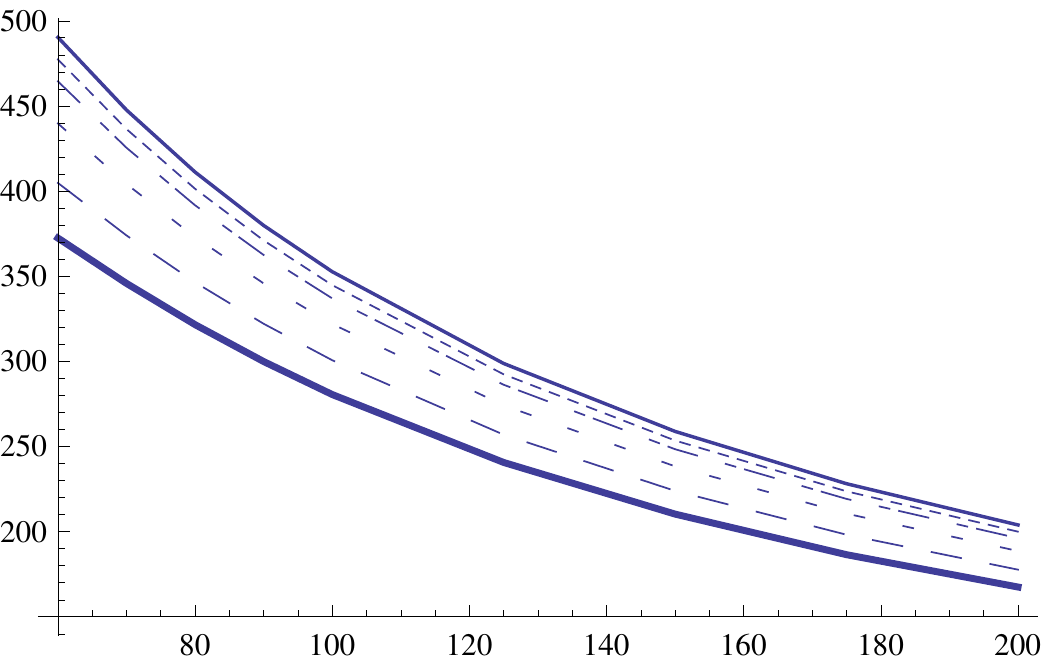}
}
\\
{\hspace{-2.0cm} $m_{\chi} \rightarrow$GeV}
\caption{ The  same as in Fig. \ref{fig:plotRs131} for $^{19}$F.
 \label{fig:plotRs19}} 
\end{center}
\end{figure}

 As we have mentioned in the case of Majorana particles only the spin induced rates become relevant. Such results for the time averaged total rate are shown in Figs \ref{fig:plotRs131} -\ref{fig:plotRs19} .
\section{Discussion}
It is clear that, for the spin contribution, the target $^{19}$F is the most favorable from an experimental point of view. This is due to the fact that the nuclear spin ME are large and the most  reliably calculated. Such large rates favor the efforts of  the PICASSO \cite{PICASSO09,PICASSO11,PICASSO12} collaboration. As we have mentioned above our nucleon cross sections are consistent with the limits placed by KIMS data \cite{KIMS12} and the earlier ZEPLIN data \cite{ZEPLIN09}.  Thus we can conclude that relatively large rates predicted for these isotops are not ruled out so far.  We have not yet seen an analysis of XENON100 \cite{XENON100.11} in terms of spin induced cross sections. Our rates may seem a bit large, but in  such an analysis it should be kept in mind that only 0.203 of the target is A=131. One has to also consider the  $A=129$ isotope. For the convenience of the experimental analysis we present the multiplicative factors needed to include to extend the data presented above in other isotopes and spin matrix elements in table \ref{tab:factor}.
\begin{table}[!t]
\caption{The multiplicative factor needed to go from the rates presented in this work to other possibilities of interest. These factors include the fraction of the isotope in question and the various spin matrix elements.}
 
\label{tab:factor}
\begin{center}
\begin{tabular}{||l|c|c|c||}
\hline
Nuclear ME&$^{129}$Xe$(100\%)$&$^{131}$Xe$(100\%)$&$^{129}$Xe$(26.4\%)$\, ,$^{131}$Xe$(21.2\%)$\\
\hline
Ressel&1&3.78&1.21\\
Finnish&0.296&2.61&0.565\\
Menendez&1.32&3.51&1.67\\
\hline
\end{tabular}
\end{center}
\end{table}
\\From table \ref{tab:factor} we see that the rates may change by factors of 4, compared to the absolute rates presented in our figures. 
 Even though the most recent calculation  \cite{MeGazSCH12} obtains results which are in agreement with the earlier  Ressel results, we should mention that
 these authors  find a retardation of the isospin=1 mode, which is the only one that contributes in our model, due to   effects of long-range two-body currents in the nucleus\cite{MeGazSCH12},\cite{MeGazSCH11}. This retardation, of about a factor of two in the rate, is essentially independent of   the nuclear target. There is also the retardation of $g_A$ inside the nuclear medium (see e.g. ref. \cite{CarHen04}) , which may lead to another factor of 2. Taking this into consideration the   rates predicted in the present work, which on the relatively large side, may be a factor of five or even 10 smaller due to uncertainties, which have nothing to do with particle model discussed here.

 In summary we have considered in this paper a viable and theoretically well motivated dark matter  candidate, which satisfies all the experimental requirements. The results presented here 
 may also apply to all models involving the WIMP  interaction with quarks via the Z-exchange.

\section*{Acknowledgments} The authors would like to thank the organizers  and participants of the Dark Side of Universe 2012 workshop for many enlightening discussions. JDV is indebted to the PICASSO collaboration for partial support of this work and to Viktor Zacek and Ubi Wichoski for their kind hospitality in Montreal and SNOLAB. This work was partially supported by UNILHC PITN-GA-2009-237920. KGS was supported in part by NSFC grant No.~10775067.

\bibliography{Tex}

\end{document}